\documentclass[12pt]{article}

\newcommand{\be}{\begin{equation}}
\newcommand{\ee}{\end{equation}}
\newcommand{\bi}[1]{\vspace{-3mm} \bibitem{#1}}

\usepackage{epsfig,amsmath,amssymb,graphics,graphicx,xcolor} 

\begin{document}

\begin{center}

{\bf \large  
On Fractional and Fractal Formulations  of \\
Gradient Linear and Nonlinear Elasticity} \\

\vskip 7mm
{\bf \large Vasily E. Tarasov} \\
\vskip 3mm

{\it Skobeltsyn Institute of Nuclear Physics,\\ 
Lomonosov Moscow State University, 
Moscow 119991, Russia} \\
{E-mail: tarasov@theory.sinp.msu.ru} \\

\vskip 7mm
{\bf \large Elias C. Aifantis}\footnote{Emeritus Professor of Engineering,
Michigan Tech, Houghton, MI 49931, USA\\
Distinguished Adjunct Professor of King Abdulaziz University,
Jeddah, 21589, SA} \\
\vskip 3mm

{\it Laboratory of Mechanics and Materials, \\ 
 Aristotle University of Thessaloniki, 
Thessaloniki 54006, Greece} \\
{E-mail: mom@mom.gen.auth.gr} \\

\vskip 7mm
\begin{abstract}
In this paper we consider extensions of 
the gradient elasticity models proposed earlier by the second 
author to describe materials with fractional non-locality and 
fractality using the techniques developed recently by the 
first author. 
We derive a generalization of 
three-dimensional continuum gradient elasticity theory, starting from integral 
relations and assuming a weak non-locality of 
power-law type that gives constitutive relations 
with fractional Laplacian terms, by utilizing 
the fractional Taylor series in wave-vector space.
In the sequel we consider non-linear field equations 
with fractional derivatives of non-integer order
to describe nonlinear elastic effects for gradient materials
with power-law long-range interactions 
in the framework of weak non-locality approximation.
The special constitutive relationship that we elaborate on,
can form the basis for developing a fractional extension of 
deformation theory of gradient plasticity. 
Using the perturbation method,
we obtain corrections to the constitutive relations
of linear fractional gradient elasticity, 
when the perturbations are caused by weak deviations 
from linear elasticity or by fractional gradient non-locality. 
Finally we discuss fractal materials described by continuum 
models with non-integer dimensional spaces. Using the recently 
suggested vector calculus for non-integer dimensional spaces,
we consider problems of fractal gradient elasticity.
\end{abstract}

\end{center}

\vskip 3mm
\noindent
PACS: 45.10.Hj; 62.20.Dc; 81.40.Jj  


\newpage
\section{Introduction}

Three-dimensional integral elasticity models for strong non-locality
are usually defined by the integral linear constitutive relation 
for the stress $\sigma_{ij}$ in terms of the strain $\varepsilon_{ij}$
in the form \cite{Kroner}--\cite{Eringen2002} 

\be \label{int-cr}
\sigma_{ij} ({\bf r},t) =
C_{ijkl} \, \varepsilon_{kl} ({\bf r},t) +
\int_{\mathbb{R}^3} c_{ijkl} ({\bf r}-{\bf r}^{\prime}) \, \varepsilon_{kl} ({\bf r}^{\prime},t) 
\, d{\bf r}^{\prime} ,
\ee

where $C_{ijkl}$ is the local fourth-order elastic 
stiffness tensor and 
$c_{ijkl} ({\bf r}-{\bf r}^{\prime})$ 
is the nonlocal elastic stiffness tensor kernel
that characterizes nonlocality of materials. 

In \cite{Aifantis1992}--\cite{Aifantis2003} it was suggested a generalization
of the constitutive relations {classical elasticity} by 
a gradient modification that contains the Laplacian $\Delta$. 
{It reads}

\be \label{H-1n}
\sigma_{ij} = C_{ijkl} \, \varepsilon_{kl} 
- l^2_s \, C_{ijkl} \, \Delta \, \varepsilon_{kl}  .
\ee

where {the newly appearing quantity}
$l_s$ is an internal length scale parameter. 
For $l_s=0$, we have the classical case of Hooke's law.
The {Lagrangian} term in equation (\ref{H-1n}) 
{-- also emerging from} the positive-definite
{strain} energy density {and a corresponding variational formulation 
\cite{AA1997} resulted to stable and well-posed solutions of boundary value 
problems and enabled the removal of singularities from dislocations and cracks 
\cite{AA2011}--\cite{AifantisILG}}. 

To describe complex materials characterized by 
non-locality of power-law type  and fractality, 
we should further generalize the gradient equation (\ref{H-1n}), 
{as already suggested by the authors \cite{TA2014},\cite{TA2015}}.

Derivatives and integrals of non-integer orders
\cite{SKM,KST} have a wide application in mechanics 
(for example see \cite{CM}--\cite{Povstenko2015}). 
The theory of fractional differential equations 
is powerful tool to describe materials and media
with power-law non-locality, 
long-range memory and/or fractality.
The fractional calculus can, in fact, be used to formulate 
a generalization of non-local theory of elasticity in both forms:
fractional gradient elasticity  
(weak power-law non-locality) and 
fractional integral elasticity 
(strong power-law non-locality).

In this paper, we consider fractional and fractal generalizations 
of gradient elasticity and plasticity models,
including {the} rather popular GRADELA model,
{as proposed and utilized by Aifantis and co-workers 
	(see, for example \cite{AA2011}--\cite{AifantisILG}, as well as the
	references quoted therein), and focus on three cases : }

\begin{itemize}
	\item The elasticity of linear materials with power-law non-locality 
	that can be described by fractional Laplacians
	of non-integer order of the Riesz type.
	\item  The elasticity of {nonlinear} materials with power-law 
	nonlocality that can be described by 
	fractional {Laplacians} of the Riesz and {Caputo} type.
	\item  The elasticity of materials with fractal structure that
	can be described in the framework of continuum models 
	by using the recently suggested vector calculus for non-integer 
	dimensional space.
\end{itemize}


{We elaborate, in particular, on} the following non-standard generalizations 
of 
gradient stress-strain relation.

(a) The fractional gradient elasticity  
with power-law non-locality  
\be \label{Gen-1a}
\sigma_{ij} = C_{ijkl} \, \varepsilon_{kl} 
- l^2_s (\alpha) \, C_{ijkl} \, (- \, ^R\Delta)^{\alpha/2} 
\, \varepsilon_{kl}  
\ee
where $(- \, ^R\Delta)^{\alpha/2}$ is the fractional generalization of the 
Laplacian in the Riesz form \cite{KST}. 

(b) The gradient elasticity for fractal materials 
\be \label{Gen-3}
\sigma_{ij} = C_{ijkl} \, \varepsilon_{kl} 
- l^2_F (D,d) \, C_{ijkl} \, \Delta^{(D,d)} \, \varepsilon_{kl}  
\ee
where $\Delta^{(D,d)}$ is the "fractal-Laplacian" 
\cite{CNSNS2015}
that takes into account the power-law density of states 
of the fractal medium under consideration.


{A new method is proposed for deriving} 
three-dimensional constitutive relation for 
fractional gradient elasticity {starting from}
the integral relations for a nonlocal elastic continuum 
{and} using the fractional Taylor series in wave-vector space.
It is well-known that constitutive relations for
gradient models can be derived from relations 
for integral nonlocal models by using the Taylor series
approximation in the coordinate space 
(see, for example, \cite{Aifantis2003}).
Specific properties of derivatives of non-integer order
do not allow us to use a fractional Taylor series
for a direct generalization of {this} approach.
{We thus} suggest to use a fractional Taylor series 
{expansion in the} wave-vector space instead of the coordinate space.
This avoids the difficulties arising from the application of
the fractional Taylor series in coordinate space. 
The physical basis of the proposed method is
an assumption {pertaining to} fractional spatial dispersion 
for a {nonlocal elastic} continuum.


{Fractional nonlinear elasticity is considered, by elaborating on}
{model non-linear field equations} 
with fractional derivatives of non-integer order
to describe materials with power-law non-locality
and weak non-linearity.
{A special constitutive equation (involving the second invariants of
deviatoric stress and strain) is employed which can be used as a starting
point for developing a fractional theory of gradient plasticity}.
Using the perturbation method, 
we obtain corrections to linear constitutive relations,
where the perturbations are caused 
by weak deviations from elasticity.
Perturbations {caused} by 
fractional gradient nonlocality are also discussed.


Fractal media and materials {in non-Euclidean space} can be described
by different methods (see, for example, \cite{Enc2018} and references
{quoted} therein).Fractal materials can be considered as 
special continua with non-integer physical dimensions (\cite{TarasovSpringer}).
Continuum models {for} fractal materials {can be} formulated using the 
fractional integration of non-integer orders. 
The kernels of fractional integrations are interpreted as
power-law densities of states, and orders of fractional integrals 
are equal to the physical dimensions of {the} material.
In these models, the {concept} of density of states is applied in addition
to the notion of distribution functions (such as 
density of mass), {thus allowing to}  take into account the specific 
properties of fractal materials. {Their interesting feature is that while}
they use fractional integration of non-integer
order, the differential operators {that include the density of states}
are of integer-order. 
{Such} models have been considered {earlier} by the {second author}
\cite{PLA2005-1}--\cite{IJMPB2005-2}, {and more recently by other authors}
\cite{MOS-01}--\cite{MOS-4}.

Fractal materials can also be described by using continuum models
with non-integer dimensional spaces (NIDS) \cite{Enc2018}, 
where integration and differentiation for NIDS are applied. 
The integration in non-integer dimensional space
is well developed \cite{Collins}--\cite{Stillinger}, and
it has a wide application in quantum field theory.
Differentiation in non-integer dimensional space is
considered in \cite{Stillinger}-\cite{PS2004}, {but in these papers} 
only a scalar Laplacian for NIDS {was suggested}.
Recently a generalization of
differential operators of first order 
(gradient, divergence, curl operators) 
and the vector Laplacian has been proposed 
in \cite{CNSNS2015,CSF2014}.
The suggested vector calculus for NIDS allows us 
to expand the range of applications
of continuum models of isotropic fractal materials. 
Generalizations of the gradient, divergence, 
curl operators and the vector Laplace operator 
for non-integer dimensional and fractional spaces 
to describe anisotropic fractal materials
have been suggested in \cite{JMP2014}.
Using the vector calculus for non-integer dimensional spaces,
which is suggested in
\cite{CNSNS2015,WM2016} to describe 
different properties of isotropic fractal media, 
it is possible to consider different problems
of fractal gradient elasticity.
The corresponding {governing} equations are differential 
equations with integer-order derivatives.
Therefore the suggested vector calculus can allow us
easy to solve the cylindrical and spherical 
(boundary value) problems
for gradient elasticity theory of fractal materials
by using tools of integer-order differential equations.
The definitions of vector operators for
non-integer dimensional spaces, can be realized for
two cases: $d=D-1$  and $d \ne D-1$,
where $D$ is the dimension of the considered fractal 
material region and $d$ is the dimension of it's boundary.

{The plan of the papers is as follows :} In Section 2 we give 
a rigorous derivation of fractional gradient elasticity, starting from 
an integral type of constitutive law of nonlocal elasticity and adopting a 
fractional Taylor series expansion for its kernel in wave-vector space. 
This expansion generates fractional Laplacians of the Riesz type and the 
resulting fractional GRADELA model is solved by the Green's function and 
Fourier transform techniques.
In Section 3, nonlinear elasticity effects are considered and a fractional 
constitutive equation involving fractional Laplacians of the Riesz and Caputo 
type is proposed. This nonlinear fractional constitutive equation can be 
utilized for constructing in the future extended fractional models of the 
standard deformation theory of plasticity. It is further used here to consider 
perturbations due to nonlinearity and fractional nonlocality.
Finally in Section 4 we consider fractal gradient elasticity based on 
Laplacians for non-integer dimensional space, generalizing the usual Laplacian 
in Euclidean space.
Various GRADELA models for fractal media are proposed and explicit form of the 
governing differential equations are derived for problems of radial symmetry. 
It is interesting that these equations for fractal media and Laplacians for 
non-integer dimensional space involve derivatives of integer order and, thus, 
they can be solved by resorting to existing methods.


\newpage

\section{Fractional Gradient Elasticity as an approximation of Nonlocal 
Elasticity}


\subsection{Derivation of general constitutive relation
for weak nonlocal elasticity}

In this section we obtain constitutive relations
for fractional gradient elasticity
from integral constitutive relations for nonlocal materials
by using a fractional Taylor series approximation
in wave-vector space. 
We start from {an} integral constitutive relation
{between stress and strain} that is a convolution 
in coordinate space.
Then we {apply} a Fourier transform of the convolution
to obtain a multiplication in the dual space 
(the wave-vector space).
Using an assumption of weak power-law 
(fractional) non-locality, which gives a spatial dispersion,
we apply a fractional Taylor series in the dual space
to obtain an adequate approximation.
Then we realize an inverse Fourier transform
that gives a constitutive relation
with {a} fractional Laplacian.
As a result, we derive a fractional gradient 
constitutive relation with a fractional generalization 
of the Laplacian in the Riesz form in coordinate space.
The main idea of the suggested approach
is to use a Taylor series in wave-vector space
instead of Taylor series in coordinate space 
that {is} usually used.
It allows us to avoid three problems
that can appear if we use the fractional Taylor series 
in coordinate space:
(a) problems with an integration of fractional Taylor series 
in coordinate space;
(b) problems with {an} exact form of kernels 
for fractional nonlocality, 
as derived by different lattice model, 
{in contrast to the suggested phenomenological approach which is free from
atomisitc details}, allows us to work in the
framework of the macroscopic approach;
(c) problems {with} derivations {arriving at an undesirable}
sign in front of the Laplacian term, {in contrast to the} 
proposed method, {which is free from sign constraints} 
in front of the Laplacian.


In the three-dimensional theory of nonlocal elasticity 
\cite{Eringen2002}, 
the nonlocal stress tensor $\sigma_{ij}$  is defined by
\be \label{N1a}
\sigma_{ij} ({\bf r},t) =
\int_{\mathbb{R}^3} {\cal K}(|{\bf r}-{\bf r}^{\prime}|) \, 
\sigma_{ij}^0 ({\bf r}^{\prime},t)  \, d{\bf r}^{\prime} ,
\ee
where ${\cal K}(|{\bf r}-{\bf r}^{\prime}|)$ 
is the interaction kernel that characterize nonlocality, and 
$\sigma_{ij}^0$  is the stress tensor of classical (local)
elasticity defined as
\be \label{sigma-0}
\sigma_{ij}^0 ({\bf r}^{\prime},t) = C_{ijkl} \,
\varepsilon_{ij} ({\bf r}^{\prime},t) ,
\ee
where $\varepsilon_{ij} ({\bf r}^{\prime},t)$
is the classical strain, and 
$C_{ijkl}$ is the fourth-order elastic stiffness tensor. 
For isotropic materials, we have 
\be \label{C-ijkl}
C_{ijkl} = \lambda \, \delta_{ij} \delta_{kl} +
\mu \, (\delta_{ik} \delta_{jl} +\delta_{il} \delta_{jk} ) , 
\ee
where $\lambda$ and $\mu$ are the Lam\'e {constants}.

Substitution of (\ref{sigma-0}) into (\ref{N1a}) gives
the nonlocal linear constitutive relation in the form
\be \label{N1}
\sigma_{ij} ({\bf r},t) =\int_{\mathbb{R}^3} {\cal K}(|{\bf r}-{\bf r}^{\prime}|) \, 
C_{ijkl} \, \varepsilon_{kl} ({\bf r}^{\prime},t) \,
d{\bf r}^{\prime} .
\ee
It is easy to see that equation (\ref{N1}) can be
derived from relations (\ref{int-cr}) with
\be
c_{ijkl} ({\bf r}-{\bf r}^{\prime}) =
C_{ijkl} \, \Bigl( {\cal K}(|{\bf r}-{\bf r}^{\prime}|) - 
\delta^{3} ({\bf r}-{\bf r}^{\prime}) \Bigr) . 
\ee
Mathematically, equation (\ref{N1}) can be considered 
as a convolution of 
the interaction kernel ${\cal K}(|{\bf r}|)$ and the strain 
tensor $\varepsilon_{kl}$ in the form
\be \label{N2}
\sigma_{ij} ({\bf r},t) = C_{ijkl} \, \Bigl( {\cal K} \, * \, \varepsilon_{kl} \Bigr) ({\bf r},t)  .
\ee
To use the Fourier method, we introduce 
the Fourier transform ${\cal F}$ -- {designated by a superimposed bar 
($\sim$)} -- as
\be \label{calF}
\tilde{\sigma}_{ij}  ({\bf k},t) = \int_{\mathbb{R}^3} \,
e^{- i({\bf k},{\bf r})} \,\sigma_{ij} ({\bf r},t) \, d{\bf r} ,
\ee
where ${\bf k}$ is the wave vector.
{Similar} formulas also hold for the strain tensor
and the interaction kernel : 
\be \label{calF-2}
\tilde{\sigma}_{ij} ({\bf k},t) = \int_{\mathbb{R}^3} \,
e^{- i({\bf k},{\bf r})} \,\varepsilon_{ij} ({\bf r},t) \, d{\bf r} ,
\ee
\be \label{calF-3}
\tilde{{\cal K}} (|{\bf k}|) = \int_{\mathbb{R}^3} \,
e^{- i({\bf k},{\bf r})} \, {\cal K} (|{\bf r}|) \, d{\bf r} .
\ee
Applying ${\cal F}$ to the convolution equation (\ref{N2}), 
we obtain 
\be \label{N3}
\tilde{\sigma}_{ij} ({\bf k},t) = C_{ijkl} \, 
\tilde{{\cal K}} (|{\bf k}|)  \, \tilde{\varepsilon}_{kl} ({\bf k},t)  .
\ee
{which can} be considered as a {general} 
constitutive relation in the wave-vector space
for nonlocal {elasticity}.


{The particular} dependence of the function $\tilde{{\cal K}} (|{\bf k}|)$
on the wave-vector $k=|{\bf k}|$ defines the type of 
spatial dispersion and it characterizes the type of
material non-locality.
For a wide class of nonlocal materials,
the wavelength $\lambda$ {satisfies} the relation
$k R_0 \sim R_0 / \lambda \ll 1$, where $R_0$ {denotes}
the characteristic length of interaction 
that usually {is of} the order of the interparticle distance.
In this case, the spatial dispersion of the elastic 
material is weak. 
To describe materials with such property 
it is {sufficient} to know the dependence of 
the function $\tilde{{\cal K}}(|{\bf k}|)$ only 
for small values $k=|{\bf k}|$, and {then} we can replace 
this function by {its} Taylor's polynomial series {expansion}.
For an isotropic continuum, we have 
\be \label{Tay1}
\tilde{{\cal K}}(|{\bf k}|) = 
\tilde{{\cal K}}(0) + a_1 \, k + a_2 \, k^2 + o(k^2) ,
\ee
where frequency {dependent} dispersion {is negleted} for simplicity, i.e.,
the {coefficients} $\tilde{{\cal K}}(0)$, $a_1$, $a_2$ 
do not depend on the frequency $\omega$.


The weak spatial dispersion of materials with {fractal}
power-law type of non-locality cannot be described by
the usual Taylor approximation. 
The fractional Taylor series is very useful for the approximation 
of non-integer power-law functions \cite{CNSNS2016-2}. 
For example, the usual Taylor series for the power-law function
\be \label{ve}
\tilde{{\cal K}}(k) = a_0 + a_{\alpha} \, k^{\alpha}  
\ee
{contains} infinite many terms for non-integer $\alpha$, while
the fractional Taylor series
of order $\alpha$ has a finite number of terms.
For example, we can consider the fractional Taylor series 
in the Odibat-Shawagfeh form \cite{OdbSh} that contains 
the Caputo fractional derivative $_0^CD^{\alpha}_k$, 
where $\alpha$ is the fractional order of differentiation, $0<\alpha<1$. 
This fractional Taylor series has the form
\be \label{Tay-Cap}
\tilde{{\cal K}}(k) = \tilde{{\cal K}}(0)
+ \frac{(\, _0^CD^{\alpha}_{k} \tilde{{\cal K}})(0)}{\Gamma (\alpha+1)} 
\, k^{\alpha}
+ \frac{(\, _0^CD^{\alpha}_{k} \, 
_0^CD^{\alpha}_{k} \tilde{{\cal K}})(0)}{\Gamma (2\alpha+1)} \,
k^{2 \alpha}+ o(k^{2 \alpha}) ,
\ee
where $\, _0^CD^{\alpha}_{k}$ is the Caputo fractional derivative 
\cite{KST} of order $\alpha$ with respect to $k=|{\bf k}|$ 
that is defined by
\be
(\, _0^CD^{\alpha}_{k} \tilde{{\cal K}})(k) = 
\left( _0I^{n-\alpha}_{k} \left( \frac{d}{dk}\right)^n \tilde{{\cal K}} \right)(k) .
\ee
Here $_0I^{\alpha}_k$ is the left-sided Riemann-Liouville fractional integral of order $\alpha >0$ 
with respect to $k$ of the form
\be \label{RLI}
(_0I^{\alpha}_{k}  \tilde{{\cal K}})(k) = \frac{1}{\Gamma(\alpha)} 
\int^k_0 \frac{  \tilde{{\cal K}}(k^{\prime}) \, dk^{\prime}}{(k-k^{\prime})^{1-\alpha}} , \quad (k>0) .
\ee
Note that the Caputo fractional derivative of a constant 
is zero like for the integer order derivative. 
In general, the third term in (\ref{Tay-Cap}), which contains
repeated fractional derivatives of order $\alpha$, 
is not the same  as the fractional derivative 
of order $2 \alpha$,
\[ (\, _0^CD^{\alpha}_k \, _0^CD^{\alpha}_k \tilde{{\cal K}})(k) \ne (\, _0^CD^{2\alpha}_k \tilde{{\cal K}})(k) . \]
The coefficients of the fractional Taylor series 
in the Odibat-Shawagfeh form 
can be found in the usual way by repeated differentiation.
Using the equation
\be \label{T2}
_0^CD^{\alpha}_k k^{\beta} = 
\frac{\Gamma(\beta+1)}{\Gamma(\beta-\alpha+1)} \, 
k^{\beta-\alpha} , \quad (k>0, \ \alpha>0, \ \beta >0)
\ee
for the case $\beta=\alpha$, in the form
\be
\, _0^CD^{\alpha}_k k^{\alpha} = \Gamma(\alpha+1), \quad (\, _0^CD^{\alpha}_k )^n k^{\alpha} =0 ,  \quad (n \ge 2) ,
\ee
we {obtain}
\be
(\, _0^CD^{\alpha}_k \tilde{{\cal K}})(0) = \Gamma(\alpha+1), \quad
( (\, _0^CD^{\alpha}_k)^n \tilde{{\cal K}})(0) = 0, \quad (n \ge 2) .
\ee
It is easy to see that the fractional Taylor series approximation of the 
function {given by equation} (\ref{ve}) is exact.


The fractional Taylor series expansion
for the function $\tilde{{\cal K}}(|{\bf k}|)$, 
can  be {written} in the form
\be
\tilde{{\cal K}}(|{\bf k}|)= a_0 + 
\sum^N_{j=1} a_{\alpha_j} |{\bf k}|^{\alpha_j} 
+ o(|{\bf k}|^{\alpha_N}) ,
\ee
where $0<\alpha_1< \alpha_2 < \ . \ . \ . \ < \alpha_N$; {with}
the {small}-o notation $o(|{\bf k}|^s)$ meaning {as usual}, 
{inclusion of} terms {with} higher powers of $|{\bf k}|$ 
than $|{\bf k}|^s$. 
The coefficients $a_{\alpha_j}$ in the Odibat-Shawagfeh form 
of the fractional Taylor series are defined by
\be \label{a-alpha-j}
a_0 = \tilde{{\cal K}}(0) , \quad
a_{\alpha_j} = \frac{1}{\Gamma (j\alpha+1)} \, 
((\, _0^CD^{\alpha}_{k} \, )^j \tilde{{\cal K}})(0) .
\ee
As a result, {for long wavelengths} we can use the approximation 
\be \label{approx-1}
\tilde{{\cal K}}(|{\bf k}|) \approx a_0 + 
\sum^N_{j=1} a_{\alpha_j} |{\bf k}|^{\alpha_j} 
\ee
for materials with power-law type of nonlocality.
The order of the fractional Taylor series approximation 
should be correlated with the type of 
power-law nonlocality 
in order {that} the fractional Taylor approximation 
of $\tilde{{\cal K}}(|{\bf k}|)$ {to be valid}. 
In the general case $0<\alpha_{j+1}- \alpha_j<1$,  
we can use the fractional Taylor formula 
in the Dzherbashyan-Nersesian form \cite{Arm1,Arm2}, 
\be \label{Arm}
\tilde{{\cal K}} (k) = \sum^{N}_{j=0} a_{\alpha_j}  
\, k^{\alpha_j} + {\cal R}_{N+1}(k), \quad (k>0) ,
\ee
where
\be \label{a-alpha-j2}
a_{\alpha_j}  = \frac{(D^{(\alpha_j)}  \tilde{{\cal K}})(0)}{
\Gamma(\alpha_j+1)} , \quad
{\cal R}_{N+1}(k) = \frac{1}{\Gamma(\alpha_{N+1}+1)} 
\int^k_0 (k-z)^{\alpha_{N+1}-1} \, 
(D^{(\alpha_j)}  \tilde{{\cal K}})(z) \, dz ,
\ee
and $\alpha_j$ $(j=0,1,...,m)$ is {an} increasing sequence of real numbers 
such that
\be
0 < \alpha_j-\alpha_{j-1} \le 1, \quad \alpha_0=0, \quad j=1,2,...,N+1 .
\ee
In equation (\ref{a-alpha-j2}) 
we use the notation \cite{Arm1,Arm2} 
(see also Section 2.8 in \cite{SKM}) of the form
\be
D^{(\alpha_j)}  = \, _0I^{1-(\alpha_j-\alpha_{j-1} )} 
\, _0D^{1+ \alpha_{j-1} }_k .
\ee
In general, {the} fractional derivative 
$D^{(\alpha_k)}$ differs from the Riemann-Liouville derivative 
\cite{SKM,KST} by a finite sum of power functions since 
$ _0I^{\alpha}_k \, _0I^{\beta}_k \ne \, _0I^{\alpha+\beta}_k$
(see Eq. 2.68 in \cite{KST}). 

For the special cases $\alpha_j= j \, \alpha$, 
where $\alpha<1$ and/or $\alpha_j=\alpha+j$, 
we could use the fractional Taylor 
formulas in the Riemann formal form  \cite{Riem,Hardy},
in the Riemann-Liouville form
(see Chapter 1. Section 2.6 \cite{SKM}),
and the Trujillo-Rivero-Bonilla form \cite{TRB}.
{In this connection, it is noted}
that the fractional Taylor series with
Caputo derivatives is {physically more meaningful} than a series 
with Riemann-Liouville derivatives, 
since the Riemann-Liouville derivative
of a constant is not equal to zero.
If $\alpha_j=j$ for all $j \in \mathbb{N}$, 
we can use the usual Taylor's formula.



\subsection{Derivation of the fractional GRADELA constitutive relation}

Let us obtain a constitutive relation
for {a special class of fractional gradient elasticity - The fractional 
GRADELA model} :

Substitution of (\ref{approx-1}) into (\ref{N3}) gives
\be \label{N3new1}
\tilde{\sigma}_{ij} ({\bf k},t) = C_{ijkl} \, 
\Bigl( a_0 + 
\sum^N_{j=1} a_{\alpha_j} |{\bf k}|^{\alpha_j} 
\Bigr) \, \tilde{\varepsilon}_{kl} ({\bf k},t)  .
\ee
where the coefficients $a_{\alpha_j}$ are defined by 
the fractional derivatives of $\tilde{{\cal K}}(|{\bf k}|)$ 
with respect to $k$ of order $\alpha_j$ at zero $k=0$.
For example, if we use
the fractional Taylor series in 
the Odibat-Shawagfeh form, then 
we should use (\ref{a-alpha-j})
\be \label{a-alpha-j3}
a_{\alpha_j} =
\frac{( (\, _0^CD^{\alpha_{j}})^j \, \tilde{{\cal K}})(0)}{\Gamma (j\alpha+1)} .
\ee
The type of the fractional Taylor series should be
correlated with the type of non-locality of {the under consideration} 
material.
In particular the first non-zero coefficient $a_{\alpha_j}$ 
is the term, $\tilde{{\cal K}}(k) - \tilde{{\cal K}}(0)$
{which is} asymptotically equivalent to $k^{\alpha_j}$ as $k \to 0$.


The inverse Fourier transform of (\ref{N3new1}) gives
the constitutive relation for fractional gradient 
elasticity in the form
\be \label{N3new3}
\sigma_{ij} ({\bf r},t) = C_{ijkl} \, 
\Bigl( \tilde{{\cal K}}(0) + \sum^N_{j=1} a_{\alpha_j} \,
(-\Delta)^{\alpha_j/2} \Bigr) \, 
\varepsilon_{kl} ({\bf r},t) ,
\ee
where we used the connection between the Riesz fractional 
Laplacian $(-\Delta)^{\alpha/2}$ 
and its Fourier transform \cite{SKM,KST},  
\be \label{FFL}
{\cal F}[ (-\Delta)^{\alpha/2} 
\varepsilon_{kl}({\bf r},t)]({\bf k})= 
|{\bf k}|^{\alpha} \, \tilde{\varepsilon}_{kl}({\bf k},t)
\ee
in the form
\be 
|k|^{\alpha_j} \longleftrightarrow (-\Delta)^{\alpha_j/2} . 
\ee

{In view of the above, we first derive the standard non-fractional}
constitutive relation for the {GRADELA model}.
We consider the special case $\alpha_j=j$ 
for integer $j \in \mathbb{N}$, 
and the function $\tilde{{\cal K}}(k)$ in the form
\be \label{a2}
\tilde{{\cal K}}(k) \approx a_0 + a_2 \, k^2 ,
\ee
where $a_{2} \ne 0$, {with} all other $a_j=0$.
Then, the inverse Fourier transform of (\ref{N3}) with (\ref{a2})
gives the constitutive relation
\be \label{N3new2}
\sigma_{ij} ({\bf r},t) = C_{ijkl} \, 
\Bigl( a_0 - a_{2} \Delta \Bigr) \, 
\varepsilon_{kl} ({\bf r},t) ,
\ee
where
\be
a_2 = \left( \frac{\partial^2 \tilde{{\cal K}}(k)}{\partial k^2} \right)_{k=0} . 
\ee
{It is also assumed that there is no initial stress, so that}
\[ a_0=\tilde{{\cal K}}(0)=1 \] 
in {consistency with the usual} the Hooke's law
\[ 
\sigma_{ij} ({\bf r},t) = C_{ijkl} \, 
\varepsilon_{kl} ({\bf r},t)  .
\] 
Let us introduce the {internal length} scale parameter $l^2_s$ 
of gradient elasticity {given by}
\be
l^2_s= |a_2| = \left| \frac{\partial^2 \tilde{{\cal K}}(k)}{\partial k^2} \right|_{k=0} . 
\ee
Then, the second-gradient term is preceded by 
a sign that is defined by the sign of $a_2$.
As a result, relation (\ref{N3new2}) 
can be rewritten in the form
\be \label{N3new2b-}
\sigma_{ij} ({\bf r},t) = C_{ijkl} \, 
\Bigl( 1 - l^2_s \, \Delta \Bigr) \, 
\varepsilon_{kl} ({\bf r},t) , \quad (for \quad a_2>0) ,
\ee
\be \label{N3new2b+}
\sigma_{ij} ({\bf r},t) = C_{ijkl} \, 
\Bigl( 1 - \operatorname{sgn}(a_{2}) \, l^2_s \, 
\Delta \Bigr) \, \varepsilon_{kl} ({\bf r},t) , 
\quad (for \quad a_2<0).
\ee
This is the well-known constitutive relation 
\cite{Aifantis1992}--\cite{AifantisILG} 
for gradient elasticity. 
The suggested approach to obtain 
constitutive relation can be generalized 
for the case of the higher order gradient elasticity 
by using {additional} integer non-zero values of $a_j$.

{Next, we extend this approach to derive the fractional counterpart of the 
GRADELA model}. We consider the case $\alpha_j=\alpha$ for some $j=j_0$, 
and $\alpha_j=0$ for all other values of $j \in \mathbb{N}$. 
Then the function $\tilde{{\cal K}}(k)$ has the form
\be \label{a2b}
\tilde{{\cal K}}(k) \approx 1 + a_{\alpha} \, k^{\alpha} .
\ee
Substitution (\ref{a2b}) into (\ref{N3}), and {subsequent application of}
the inverse Fourier transform gives the constitutive relation
\be \label{N3new3c}
\sigma_{ij} ({\bf r},t) = C_{ijkl} \, 
\Bigl( 1 +  a_{\alpha} \,
(-\Delta)^{\alpha/2} \Bigr) \, 
\varepsilon_{kl} ({\bf r},t) .
\ee
Using the new scale parameter $l_s(\alpha)$, equation (\ref{N3new3c})
can be written in the form
\be \label{N3new3d-}
\sigma_{ij} ({\bf r},t) = C_{ijkl} \, 
\Bigl( 1 + l^2_s (\alpha) \, (-\Delta)^{\alpha/2} \Bigr) \, 
\varepsilon_{kl} ({\bf r},t) , \quad (for \quad a_{\alpha} >0 ) ,
\ee
\be \label{N3new3d+}
\sigma_{ij} ({\bf r},t) = C_{ijkl} \, 
\Bigl( 1 - l^2_s (\alpha) 
\, (-\Delta)^{\alpha/2} \Bigr) \, \varepsilon_{kl} ({\bf r},t) , 
\quad (for \quad a_{\alpha} < 0 ) ,
\ee 
where $l^2_s(\alpha)=|a_{\alpha}|$ is the scale parameter.
For $\alpha=2$, the 
relation (\ref{N3new3d-}) gives (\ref{N3new2b-}),  
and the relation (\ref{N3new3d+}) gives (\ref{N3new2b+}). 
If we use the fractional Taylor series in 
the Odibat-Shawagfeh form (\ref{a-alpha-j3}), then
the scale parameter is defined by
\be \label{l2s-alpha}
l^2_s (\alpha) = |a_{\alpha}| =  
\frac{1}{\Gamma (j\alpha+1)} \, 
\left| \Bigl( (\, _0^CD^{\alpha/j}_{k} )^j \tilde{{\cal K}}\Bigr)(0) \right| .
\ee 
In general, for the Caputo fractional derivative
in equation (\ref{l2s-alpha}) we have the inequality
\be
(\, _0^CD^{\alpha/j}_{k} )^j \ne \, _0^CD^{\alpha}_{k} .
\ee  
For example, for $j=2$ we have
\be \label{VSP}
\Bigl((\, _0^CD^{\alpha/2}_k )^2 \tilde{{\cal K}}\Bigr)(0)  =
(\, _0^CD^{\alpha}_k \tilde{{\cal K}})(0)  +
\frac{ (D^1_k\tilde{{\cal K}})(0) }{\Gamma(1-\alpha)} 
\, k^{1-\alpha} . \ee
Note that ${\bf r}$, $r$ and $l^2_s(\alpha)$ 
are dimensionless {quantities} for fractional elasticity. 
Equation (\ref{N3new3d-}) and (\ref{N3new3d+}) are the 
constitutive relations for fractional {GRADELA model}.
One of the advantages of the suggested phenomenological 
approach is its independence {on the details of the underlying 
micro/nanostructures}.
As a result, the {above} fractional constitutive relation 
can correspond to different lattice models.


{The governing equations for the components} 
$u_{i} ({\bf r},t)$ of
the displacement vector can be derived form the 
{linear momentum balance equation for continuous media}, i.e.
\be \label{law}
\rho \, \ddot{u}_i ({\bf r},t) = \sum_{j} 
\frac{\partial \sigma_{ij} ({\bf r},t)}{\partial x_j} 
+ f_i ({\bf r},t) ,
\ee
where $f_i$ {denotes body force}. The {linearized} strain tensor 
is connected {to} the displacement vector {through the usual relation}
\be \label{ekl}
\varepsilon_{kl} ({\bf r},t) = \frac{1}{2} 
\left( \frac{\partial u_{k} ({\bf r},t)}{\partial x_l} +
\frac{\partial u_{l} ({\bf r},t)}{\partial x_k} \right) .
\ee
Substitution of (\ref{ekl}) and of the constitutive 
relation (\ref{N3new3c}) for fractional gradient elasticity
into equation (\ref{law}), {results to governing} equation 
for {the} displacement {of the fractional GRADELA model},
in the form
\be \label{C-eq}
\rho \, \frac{\partial^2 u_i ({\bf r},t)}{\partial t^2} = 
\sum_{j,l,m} C_{ijlm} \,
\frac{\partial^2 u_m({\bf r},t)}{\partial x_j \partial x_l} +
\sum_{j,l,m} C_{ijlm} \; a_{\alpha} \, 
\frac{\partial}{\partial x_j} \, (-\Delta)^{\alpha/2} \,
\frac{\partial u_m({\bf r},t)}{\partial x_l}
+ f_i({\bf r},t) ,
\ee
Let us {now} consider the one-dimensional case, {with}
\be
u_x ({\bf r},t) = u (x,t) , \quad 
u_y ({\bf r},t)=u_z ({\bf r},t)=0 , 
\ee
\be
f_x ({\bf r},t) = f (x,t) , \quad 
f_y ({\bf r},t) = f_z ({\bf r},t)=0 .
\ee
In this case, $C_{111}=\lambda + 2 \, \mu$, and
equation (\ref{C-eq}) has the form
\be \label{C-eq-2}
\rho \, \frac{\partial^2 u (x,t)}{\partial t^2} = 
(\lambda + 2 \, \mu) \, \left(
\frac{\partial^2 u(x,t)}{\partial x^2} + a_{\alpha} \, 
\frac{\partial}{\partial x} 
\frac{\partial^{\alpha}}{\partial |x|^{\alpha}} 
\frac{\partial u(x,t)}{\partial x} \right) + f(x,t) ,
\ee
where $\partial^{\alpha} / \partial |x|^{\alpha}$
is the Riesz fractional derivative \cite{SKM,KST} 
with respect to $x \in \mathbb{R}$.
Note that $\partial^{\alpha} / \partial |x|^{\alpha}$
for $\alpha=2$ is a derivative of second order 
with respect $x$ with the minus sign
\be
\frac{\partial^{\alpha=2}}{\partial |x|^{\alpha=2}} =
- \frac{\partial^2}{\partial x^2} .
\ee
For {the cases of} $\alpha=1$ {and of others with} odd integer 
values, $\partial^{\alpha} / \partial |x|^{\alpha}$
cannot be considered as a local operator.
It is a nonlocal operator; e.g.
\be
\frac{\partial^{\alpha=1}}{\partial |x|^{\alpha=1}} \ne
\frac{\partial}{\partial x} .
\ee
For the static case ($u(x,t)=u(x)$, $f(x,t)=f(x)$), 
equation (\ref{C-eq-2}) can be rewritten in the form
\be \label{C-eq-3a}
\frac{\partial^2 u(x)}{\partial x^2} + a_{\alpha} \, 
\frac{\partial}{\partial x} 
\frac{\partial^{\alpha}}{\partial |x|^{\alpha}} 
\frac{\partial u(x)}{\partial x} 
= - \frac{1}{\lambda + 2 \, \mu} f(x) .
\ee
For $\alpha=2$, equation (\ref{C-eq-3a})
describes the static equation for usual (non-fractional)
gradient elasticity {in one dimension} \cite{TA2014}.


\subsection{Green functions and Fourier method 
to solve fractional differential equations}


Let us consider the fractional partial differential equation
in the form
\be \label{FPDE-1}
\sum^m_{j=1} a_j ((-\Delta)^{\alpha_j/2} \Phi) ({\bf r}) + a_0 \Phi ({\bf r})= f({\bf r}) ,
\ee   
where $\alpha_m > ... > \alpha_1>0$, and $a_j \in \mathbb{R}$ 
($1 \leq j \leq m$) are constants.
{We apply the Fourier method, which is based on the relations} 
\begin{equation} \label{FFL2}
\begin{split}
{\cal F}[ \frac{ \partial^{\alpha_j} \Phi ({\bf r}) } 
	           {  \partial | x_i |^{\alpha_j}} ] ({\bf k}) = 
	           | k_i |^{\alpha_j}  \, \hat \Phi ({\bf k}).\\
{\cal F}[ (-\Delta)^{\alpha_j/2} \Phi ({\bf r})]({\bf k})= 
|{\bf k}|^{\alpha_j} \, \hat \Phi ({\bf k}).
\end{split}
\end{equation}
{for Riesz fractional derivatives and the Riesz Laplaican, valid for
the Lizorkin space \cite{SKM} and infinitely differential functions 
$C^{\infty} ({\mathbb R^1}\!)$ on $\mathbb R^1$ with compact support.}
Applying the Fourier transform ${\cal F}$ to both sides of (\ref{FPDE-1}) and using (\ref{FFL2}), we have
\be
({\cal F} \Phi)({\bf k}) = \left( \sum^m_{j=1} a_j |{\bf k}|^{\alpha_j}+a_0 \right)^{-1} 
({\cal F} f)({\bf k}) .
\ee

The fractional analog of the Green function (see Section 5.5.1. in \cite{KST}) is given by
\be \label{FGF}
G_{\alpha}({\bf r})= {\cal F}^{-1} \Bigl[ \left( \sum^m_{j=1} a_j |{\bf k}|^{\alpha_j}+
a_0 \right)^{-1} \Bigr] ({\bf r})=
\int_{\mathbb{R}^3} \left( \sum^m_{j=1} a_j |{\bf k}|^{\alpha_j}+a_0 \right)^{-1} \
e^{ + i ({\bf k},{\bf r}) } \, d^3 {\bf k} ,
\ee
where $\alpha=(\alpha_1,...,\alpha_m)$.
The following relation
\be \label{3-1}
\int_{\mathbb{R}^n} e^{  i ({\bf k},{\bf r}) } \, G(|{\bf k}|) \, d^n {\bf k}= 
\frac{(2 \pi)^{n/2}}{ |{\bf r}|^{(n-2)/2}} 
\int^{\infty}_0 G( \lambda) \, \lambda^{n/2} \, J_{n/2-1}(\lambda |{\bf r}|) \, 
d \lambda,
\ee
holds (see Lemma 25.1 of \cite{SKM}) for 
any suitable function $G$
such that the integral in the right-hand side of (\ref{3-1}) is convergent. 
Here $J_{\nu}$ is the Bessel function of the first kind. As a result, the Fourier transform of a radial function is also a radial function.

On the other hand, using (\ref{3-1}), the Green function (\ref{FGF}) can be represented (see Theorem 5.22 in \cite{KST})
in the form of the one-dimensional integral involving the Bessel function $J_{1/2}$ of the first kind 
\be \label{G-1}
G_{\alpha} ({\bf r}) =\frac{|{\bf r}|^{-1/2}}{(2 \pi)^{3/2}} 
\int^{\infty}_0 \left( \sum^m_{j=1} a_j |\lambda|^{\alpha_j}+a_0 \right)^{-1} 
\lambda^{3/2} \, J_{1/2} (\lambda |{\bf r}|) \, d \lambda,
\ee
where we use $n=3$ and $\alpha=(\alpha_1,...,\alpha_m)$.
For {the} one-dimensional case, {we have}
\be
J_{-1/2} (z) = \sqrt{\frac{2}{\pi z}} \, \cos (z) ,
\ee
{whereas} for three-dimensional case, {we have} 
\be
J_{1/2} (z) = \sqrt{\frac{2}{\pi z}} \, \sin (z) .
\ee

If $\alpha_m > 1$ and $A_m \ne 0$, $A_0 \ne 0$, then equation (\ref{FPDE-1})  
(see, for example, Section 5.5.1. pages 341-344 in  \cite{KST}) has a 
particular solution represented in the form of the convolution of the functions 
$G({\bf r})$ and $f({\bf r})$ as follows
\be \label{phi-G}
\Phi({\bf r})=  
\int_{\mathbb{R}^3} G_{\alpha} ({\bf r} - {\bf r}^{\prime}) \, 
f ({\bf r}^{\prime}) \, d^3 {\bf r}^{\prime},
\ee
where the Green function $G_{\alpha}(z)$ is given by (\ref{G-1}).
{Thus, we can now effectively consider} the fractional partial differential 
equation 
(\ref{FPDE-1}) with $a_0=0$ and $a_1 \ne 0$, when $m \in \mathbb{N}$, $m \ge 
1$, {as well as} the case where $\alpha_1< 3$, $\alpha_m > 1$, $m \ge 1$, 
$a_1 
\ne 
0$, $a_m \ne 0$, $\alpha_m > ... > \alpha_1>0$, {i.e. the equation}
\be \label{FPDE-3}
\sum^m_{j=1} a_j ((-\Delta)^{\alpha_j/2} \Phi) ({\bf r}) = 
 f({\bf r}) .
\ee
The above equation has the following particular solution (see Theorem 5.23 in 
\cite{KST}) 
\be \label{phi-G3}
\Phi({\bf r})= 
\int_{\mathbb{R}^3} G_{\alpha} ({\bf r} - {\bf r}^{\prime}) \, 
f ({\bf r}^{\prime}) \, d^3 {\bf r}^{\prime} ,
\ee
with
\be \label{G-3}
G_{\alpha} ({\bf r}) =\frac{|{\bf r}|^{-1/2}}{(2 \pi)^{3/2}} 
\int^{\infty}_0 \left( \sum^m_{j=1} a_j |\lambda|^{\alpha_j} \right)^{-1} 
\lambda^{3/2} \, J_{1/2} (\lambda |{\bf r}|) \, d \lambda \, ,
\ee
{which also describes} the
electrostatic field in plasma-like media with 
a spatial dispersion of power-law type.


Let us now apply the multi-dimensional Fourier {method} 
to derive particular solutions for the linear fractional 
differential equations (\ref{C-eq}) by {employing} the 
direct and inverse Fourier transforms, and using the 
The Fourier transform method for solving the relations
\be
{\cal F} \left\{ 
\frac{\partial u_i({\bf r},t)}{\partial x_l}
\right\} ({\bf k},t) = 
\, i \, k_l \, {\cal F} \{ u_i({\bf r},t) \} ({\bf k},t) ,
\ee
\be
{\cal F} \left\{ (-\Delta)^{\alpha/2} \, 
u_i({\bf r},t) \right\} ({\bf k},t) = |{\bf k}|^{\alpha} \, 
{\cal F} \{ u_i({\bf r},t) \} ({\bf k},t) .
\ee
The Fourier transform of equation (\ref{C-eq}) has the form
\be \label{C-eq2}
\rho \, \frac{\partial^2 \tilde{u}_i ({\bf k},t) }{\partial t^2} = \sum_{j,l,m} C_{ijlm} \, \Bigl(
- k_j \, k_l - a_{\alpha} \, k_j \, k_l \, |{\bf k}|^{\alpha}   \Bigl) \, 
\tilde{u}_m ({\bf k},t) + \tilde{f}_i({\bf k},t) ,
\ee
where $\tilde{f}_i({\bf k},t) =
{\cal F} \{ f_i({\bf r},t) \} ({\bf k},t)$. Using the tensor
\be \label{hatC}
\hat{C}_{im} ({\bf k}) 
= \sum_{j,l} C_{ijlm} \, k_j \, k_l \, 
(1 + a_{\alpha} \, |{\bf k}|^{\alpha} ) ,
\ee
we can {calculate its} inverse 
$\hat{C}^{-1}_{im}({\bf k})$ {through the identity}
\be
\sum_m \hat{C}^{-1}_{im} ({\bf k}) \, \hat{C}_{ml} ({\bf k}) 
= \delta_{il} .
\ee

For {the} static case 
($\tilde{u}_i ({\bf k},t)=\tilde{u}_i ({\bf k})$,
$\tilde{f}_i ({\bf k},t)=\tilde{f}_i ({\bf k})$), 
equation (\ref{C-eq2}) has the form
\be
- \sum_m \hat{C}_{im} ({\bf k}) \,\tilde{u}_m ({\bf k}) + 
\tilde{f}_i ({\bf k}) = 0 ,
\ee
and {thus}
\be \label{Sol-1}
\tilde{u}_m ({\bf k}) = \sum_i
\hat{C}^{-1}_{mi} ({\bf k}) \, \tilde{f}_i({\bf k}) .
\ee
Applying the inverse Fourier transform to (\ref{Sol-1}), 
we obtain a particular solution to equation (\ref{C-eq}) 
in the form
\be \label{Sol-2}
u_m ({\bf r}) = \sum_i \, {\cal F}^{-1} \left\{  
\hat{C}^{-1}_{mi} ({\bf k}) \, 
\tilde{f}_i ({\bf k}) \right\} .
\ee
{On introducing} the fractional 
analog of the Green function
\be
G^{mi} ({\bf r}) = {\cal F}^{-1} \left\{  
\hat{C}^{-1}_{mi} ({\bf k}) \right\} .
\ee
and applying the {convolution} property of the Fourier 
transform, we can {then represent} the solution (\ref{Sol-2}) 
in the form of the convolution of the Green's function 
$G^{mi} ({\bf r})$ and the {body} force $f_i({\bf r})$ as
\be \label{Sol-3}
u_m ({\bf r}) = \int_{\mathbb{R}^3} 
G^{mi} ( {\bf r} -  {\bf r}^{\prime} ) \, 
f_i ({\bf r}^{\prime} )\, d {\bf r}^{\prime} .
\ee


Let us now obtain a solution for the  
one-dimensional equation (\ref{C-eq-3a}) {in Fourier space, i.e.}
\be
( k^2+ a_{\alpha} \, k^{\alpha+2} ) \, 
\tilde{u} (k) = \frac{1}{\lambda + 2 \, \mu} \tilde{f} (k) ,
\quad (\alpha>0) ,
\ee
where $k=k_x$, {by utilizing the Green's function} (see Theorem 5.24 in 
\cite{KST}) in the form
\be 
G^1_{\alpha} (x) =\frac{1}{\pi} 
\int^{\infty}_0 \frac{ \cos (\lambda |x|) }{ \lambda^2 + 
a_{\alpha} \, \lambda^{\alpha+2}} \, d \lambda .
\ee
For the deformation at positions $x$, which are {large in comparison} with 
the size of the region where the force is applied (point load), we can assume 
that the force is given by
\be \label{deltaf}
f(x) =f_0 \delta(x) . 
\ee
Then the {displacement}, which is a particular solution of equation 
(\ref{C-eq-3a}), will be
described by the equation
\be \label{SOL-1}
u(x) = \frac{f_0}{\pi (\lambda + 2 \, \mu)  } \int^{\infty}_0 
\frac{\cos(\lambda |x|)}{ \lambda^2 +  a_{\alpha} \lambda^{\alpha+2}} \, d\lambda .
\ee
The solution (\ref{SOL-1}) for non-integer $\alpha>0$ 
describes a solution for 
the fractional {GRADELA model}.


\subsection{Solution based on Fourier transform} 

The fractional {GADELA displacements for the static case are given by the 
equation}
\be \label{1}
\mu \, u_{i,jj} + (\lambda +\mu) \, u_{j,ji} + l^2_{\alpha} \, (-\Delta)^{\alpha/2} \,
\Bigl( \mu \, u_{i,jj} + (\lambda +\mu) \, u_{j,ji} \Bigr) + f_i =0 ,
\ee
where $(-\Delta)^{\alpha/2}$ is the fractional Laplacian of order $\alpha>0$ of 
Riesz type.
For $\alpha=1$, equation (\ref{1}) gives {the usual equations of the 
standard GRADELA model \cite{AA1997}--\cite{AifantisILG}}
\be \label{2}
\mu \, u_{i,jj} + (\lambda +\mu) \, u_{j,ji} - l^2_2 \, \Delta
\Bigl( \mu \, u_{i,jj} + (\lambda +\mu) \, u_{j,ji} \Bigr) + f_i =0 .
\ee

Using the Fourier transform
\be \label{3}
\hat{F} ({\bf k}) = \int^{+\infty}_{-\infty} F({\bf x}) \, 
e^{ -i \, ({\bf k}, {\bf x})} \, d^3{\bf x} ,
\ee
\be \label{4}
F ({\bf x}) = \frac{1}{(2 \, \pi)^3} \,
\int^{+\infty}_{-\infty} \hat{F}({\bf k}) \, e^{ i \, ({\bf k}, {\bf x})} \, d^3{\bf k} ,
\ee
we obtain
\be \label{5}
|{\bf k}|^2 \, (1+l^2_{\alpha} \, |{\bf k}|^{\alpha}) \,
\Bigl( (\lambda + 2 \, \mu) \, k^0_i \, k^0_j + 
\mu (\delta_{ij}- k^0_j \, k^0_j ) \Bigr) \, \hat{u}_i({\bf k}) 
= \hat{f}_i ({\bf k}) ,
\ee
where
\be \label{6}
|{\bf k}| = (k_i\, k_i)^{1/2} , \quad k^0_i =k_i /| {\bf k}| .
\ee
Equation (\ref{5}) leads us to the expression
\be
\hat{u}_i({\bf k}) = \hat{G}_{ij}({\bf k}) \, \hat{f}_j ({\bf k}) ,
\ee
where
\be
\hat{G}_{ij}({\bf k}) = \frac{1}{|{\bf k}|^2 \, 
(1+l^2_{\alpha} \, |{\bf k}|^{\alpha}) } \,
\left( \frac{1}{\mu}(\delta_{ij}- k^0_j \, k^0_j ) 
+ \frac{1}{\lambda+2\mu} \, k^0_i \, k^0_j \right) .
\ee
Then displacement $u_i({\bf x})$ is represented in the form
\be
u_i({\bf x}) = \int^{+\infty}_{-\infty} G_{ij} ({\bf x}-{\bf y}) \, d^3 {\bf y} .
\ee
where
\be
G_{ij}({\bf x}) = \frac{1}{(2\, \pi)^3} 
\int^{+\infty}_{-\infty} \hat{G}_{ij} ({\bf k}) \, e^{ i \, ({\bf k}, {\bf x})} \, d^3 {\bf k} .
\ee

To explicitly evaluate $G_{ij}({\bf x})$, we use spherical coordinates
$(k,\theta, \phi)$ with angle $\theta$ between ${\bf x}$ ($\theta=0$) and ${\bf k}$:
\be
({\bf k},{\bf x}) =k_i \, x_i = 
|{\bf k}| \, |{\bf x}| \, \cos \theta .
\ee
\be
x =|{\bf x}| , \quad k=|{\bf k}|, \quad
d^3{\bf k}= k^2 \, dk \, \sin \theta \, d \theta \, d \phi .
\ee
Then we have 
\[
G_{ij}({\bf x}) = \frac{1}{(2\, \pi)^3} 
\int^{2 \pi}_0 \Bigl( 
\int^{\pi}_{0} \Bigl( 
\int^{\infty}_0 \Bigl(
\frac{1}{|{\bf k}|^2 \, (1+l^2_{\alpha} \, |{\bf k}|^{\alpha}) } \,
\Bigl( \frac{1}{\mu} (\delta_{ij}- k^0_i \, k^0_j ) +
\]
\be
+ \frac{1}{\lambda+2\mu} \, k^0_i \, k^0_j \Bigr)
\, e^{ i \, k\, x\, \cos \theta} \, d k \Bigr)
\sin \theta \, d \theta \Bigr) \Bigr)
\ d \phi =
\ee
\[
= \frac{1}{(2\, \pi)^3} \int^{\pi}_0 \Bigl( 
\int^{2\pi}_{0} \left( 
\left( \frac{1}{\mu} \, (\delta_{ij}- k^0_i \, k^0_j ) 
+ \frac{1}{\lambda+2\mu} \, k^0_i \, k^0_j \right) 
\ d \phi \right) \cdot 
\]
\be \label{14}
\cdot \int^{\infty}_0 \left( 
\frac{1}{(1+l^2_{\alpha} \, |{\bf k}|^{\alpha}) } \,
e^{ i \, k \, x \, \cos \theta} \, d k \right) 
\Bigr) \sin \theta \, d \theta .
\ee

Using that $\hat{G}_{ij}(-{\bf k}) = \hat{G}_{ij}({\bf k})$
and $G_{ij}(-{\bf x}) =G_{ij}({\bf x})$
{along with} the identities
\be
\int^{2 \pi}_0 \, k^0_i \, k^0_j \, d \phi 
= \pi \, \Bigl( \delta_{ij} \, \sin^2 \theta 
- x^0_i \, x^0_j (1-3 \cos^2 \theta) \Bigr) ,
\ee
where $x^0_i =x_i / |{\bf x}|$,
the Fourier transform can be represented as 
a cosine Fourier transform
\[
\int^{\infty}_0 
\frac{1}{ 1+l^2_{\alpha} \, |{\bf k}|^{\alpha} } \,
e^{ i \, k \, x \, \cos \theta} \, d k =
\frac{1}{2} \int^{\infty}_{-\infty}
\frac{1}{ 1+l^2_{\alpha} \, |{\bf k}|^{\alpha} } \,
e^{ i \, k \, x \, \cos \theta} \, d k =
\]
\be
= \int^{\infty}_0 \frac{1}{ 1+l^2_{\alpha} \, |{\bf k}|^{\alpha} } \,
\cos ( k \, x \, \cos \theta ) \, d k .
\ee

For $\alpha=2$, we can use equation (11) of Section 1.2 of \cite{ErdelyiTable} 
in the form 
\be
\int^{\infty}_0 \frac{1}{k^2+a^2} \cos (k\, x) \, dk =
\frac{\pi}{2a} \, e^{-ax} \quad (\alpha>0) .
\ee

For $\alpha \ne 2$, the corresponding integrals
cannot be evaluated explicitly.
For this integrals we introduce the notation
\be \label{Cn}
C_{n,\alpha} (x, l) := 
\int^{\infty}_0 \frac{\cos ( k \, x )}{ k^n \, (1+ l^2_{\alpha} 
\, k^{\alpha}) } \, \, d k  \quad ( n \in \mathbb{N}).
\ee
Note that 
\be
C_{0,2} (x, 1) = \frac{\pi \, e^{-x}}{2} ,
\ee
and the integral $C_{n,2} (x, 1)$ does not converge 
for $n \in \mathbb{N}$.
Usually, the integration with respect to $k$ is realized 
{first}, and then
the integration with respect to $\theta$. {Here,}
we will use a reverse sequence of integrations.
Using the notation (\ref{Cn}), 
we rewrite equation (\ref{14}) in the form
\[
G_{ij}({\bf x}) = \frac{1}{(2\pi)^3} \int^{1}_{-1} 
\Bigl( \frac{2}{\mu} + 
\Bigl(
\frac{1}{\lambda+ 2 \mu} -\frac{1}{\mu} 
\Bigr) \, (1-t^2) 
\Bigr) \delta_{ij} \, C_{0,\alpha} (x \, t , l) \, dt -
\]
\be \label{20}
- \frac{1}{(2\pi)^3} \int^{1}_{-1} \Bigl( 
\frac{1}{\lambda+ 2 \mu} -\frac{1}{\mu} 
\Bigr) 
\, x^0_i \, x^0_j \, (1-3t^2) 
\, C_{0,\alpha} (x \, t , l)  \, dt ,
\ee
where 
\[ t = \cos \theta . \]

To explicitly evaluate this expression (\ref{20}),
we should {first calculate} the integrals
\be \label{21}
\int^{1}_{-1} C_{0,\alpha} (x \, t , l) \, dt =
\int^{\infty}_0 \frac{1}{ 1+ l^2_{\alpha} \, k^{\alpha} } 
\left( \int^{1}_{-1} \cos ( k \, x \, t ) \, dt \right) \, dk .
\ee
\be
\int^{1}_{-1} C_{0,\alpha} (x \, t , l) \, (1-t^2) dt =
\int^{\infty}_0 \frac{1}{ 1+ l^2_{\alpha} \, k^{\alpha} } 
\left( \int^{1}_{-1} \cos ( k \, x \, t ) \, (1-t^2) dt \right) \, dk .
\ee
\be \label{23}
\int^{1}_{-1} C_{0,\alpha} (x \, t , l) \, (1-3 \, t^2) dt =
\int^{\infty}_0 \frac{1}{ 1+ l^2_{\alpha} \, k^{\alpha} } 
\left( \int^{1}_{-1} \cos ( k \, \, x\, t ) \, (1- 3\, t^2) dt \right) \, dk .
\ee
The integrals, which appear in the brackets
of (\ref{21})-(\ref{23}), can be explicitly 
represented by elementary functions
\be \label{24}
\int^{1}_{-1} \cos ( k \, \, x\, t ) \, dt = 
\frac{2 \, \sin(k \, x)}{k \, x} ,
\ee
\be
\int^{1}_{-1} \cos ( k \, \, x\, t ) \, (1-t^2) \, dt = 
\frac{4 \, \sin(k \, x) - 4 \, k \, x\, \cos(k \, x)}{
(x\, k)^3} ,
\ee
\be \label{26}
\int^{1}_{-1} \cos ( k \, \, x\, t ) \, (1-3t^2) dt = 
- \frac{4 \, ((x\,k)^2-3) \, \sin(k \, x) + 
12 \, k \, x \, \cos(k \, x)}{(k \, x)^3} .
\ee
Expressions (\ref{24})-(\ref{26}) can be substituted into
(\ref{21})-(\ref{23}).
{This results to }
new special functions $A_1(\alpha;l;x)$,
$A_2(\alpha;l;x)$, and $A_3(\alpha;l;x)$,
since the corresponding integrals cannot be represented
by {elementary} and well-known special functions
These new functions are defined by the equations
\be \label{A1}
A_1(\alpha;l;x) := \int^{\infty}_0 \,
\frac{2 \, \sin(k \, x)}{k \, x \, 
(1+ l^2_{\alpha} \, k^{\alpha})} \, dk .
\ee
\be \label{A2}
A_2(\alpha;l;x) := \int^{\infty}_0 \,
\frac{4 \, \sin(k \, x) - 4 \, k \, x\, \cos(k \, x)}{
(x\, k)^3 \, (1+ l^2_{\alpha} \, k^{\alpha})} \, dk .
\ee
\be \label{A3}
A_3(\alpha;l^2;x):= -\int^{\infty}_0 \,
\frac{4 \, ((x\,k)^2-3) \, \sin(k \, x) + 
12 \, k \, x \, \cos(k \, x)}{(k \, x)^3 
\, (1+ l^2_{\alpha} \, k^{\alpha})} \, dk .
\ee

Note that for the case $\alpha=2$, 
the functions (\ref{A1})-(\ref{A3})
can be explicitly represented by elementary and 
well-known special functions. 
For example, {for} $l=1$, we have
\[
A_1(2; 1; x)= 
x^{-1}\Bigl(
{ \pi  - \pi \,{\rm cosh}(x^{3/2}) + I
\,{\rm Ci}( - I\,x^{3/2})\,{\rm sinh}({\it x
\symbol{126}}^{3/2}) - I\,{\rm Ci}(I\,x^{3/2})
\,{\rm sinh}(x^{3/2})} \Bigr) 
\]
\begin{eqnarray*}
\lefteqn{ A_2(2; 1; x)= - x^{-13/2}\Bigl(3\,\pi \,
\sqrt{x} - 2\,\pi \,\sqrt{x}
\,{\rm cosh}(x^{5/2}) + 2\,x
^{3}\,\pi \,{\rm sinh}(x^{5/2})} \\
 & & \mbox{} + 2\,I\,\sqrt{x}\,{\rm Ci}( - I\,
x^{5/2})\,{\rm sinh}(x^{5/2})
 + 2\,I\,x^{3}\,{\rm Ci}(I\,x
^{5/2})\,{\rm cosh}(x^{5/2}) \\
 & & \mbox{} - 2\,I\,\sqrt{x}\,{\rm Ci}(I\,
x^{5/2})\,{\rm sinh}(x^{5/2})
 - 2\,I\,x^{3}\,{\rm Ci}( - I\,{\it x
\symbol{126}}^{5/2})\,{\rm cosh}(x^{5/2}) \Bigr)
 .
\end{eqnarray*}
\begin{eqnarray*}
\lefteqn{ A_3(2; 1; x)= x^{-13/2} \, \Bigl( - 2\,\pi \,
x^{11/2} + 2\,\pi \,{\rm cosh}({\it x
\symbol{126}}^{5/2})\,x^{11/2} - 9\,\pi \,
\sqrt{x} + 6\,\pi \,\sqrt{x}
\,{\rm cosh}(x^{5/2})} \\
 & & \mbox{} - 6\,x^{3}\,\pi \,{\rm sinh}({\it 
x\symbol{126}}^{5/2}) - 6\,I\,x^{3}\,{\rm Ci}(I
\,x^{5/2})\,{\rm cosh}(x^{5/2
}) - 6\,I\,\sqrt{x}\,{\rm Ci}( - I\,{\it x
\symbol{126}}^{5/2})\,{\rm sinh}(x^{5/2}) \\
 & & \mbox{} - 2\,I\,{\rm Ci}( - I\,x^{5/2})\,
{\rm sinh}(x^{5/2})\,x^{11/2}
 + 6\,I\,\sqrt{x}\,{\rm Ci}(I\,{\it x
\symbol{126}}^{5/2})\,{\rm sinh}(x^{5/2}) \\
 & & \mbox{} + 2\,I\,{\rm Ci}(I\,x^{5/2})\,
{\rm sinh}(x^{5/2})\,x^{11/2}
 + 6\,I\,x^{3}\,{\rm Ci}( - I\,{\it x
\symbol{126}}^{5/2})\,{\rm cosh}(x^{5/2}) \Bigr) ,
\end{eqnarray*}
where ${\rm Si}(z)$ and ${\rm Ci}(z)$
are sine and cosine {integrals}, respectively.

Using (\ref{A1})-(\ref{A3}), equation (\ref{20}) can be 
represented in the form
\[
G_{ij}({\bf x}) = \frac{1}{(2\pi)^3} \, \Bigl( 
\frac{1}{\mu} \, A_1(\alpha;l;x)\, \delta_{ij} 
+ \Bigl( \frac{1}{\lambda+ 2 \mu} -\frac{1}{\mu} 
\Bigr) \, A_2(\alpha;l;x) \, \delta_{ij} -
\]
\be
- \Bigl( \frac{1}{\lambda+ 2 \mu} -\frac{1}{\mu} \Bigr) 
\, x^0_i \, x^0_j \, A_3(\alpha;l;x) \Bigr) .
\ee

As a result, we obtain the equation
\be
G_{ij}({\bf x}) = \frac{1}{(2\pi)^3} \, \left( 
\psi_{\alpha}(x) \, \delta_{ij} -
\chi_{\alpha} (x) \, x^0_i \, x^0_j \right) ,
\ee
where
\be
\psi_{\alpha}(x) := \frac{1}{\mu} \, A_1(\alpha;l;x)
+ \Bigl( \frac{1}{\lambda+ 2 \mu} -\frac{1}{\mu} 
\Bigr) \, A_2(\alpha;l;x) ,
\ee
\be
\chi_{\alpha} (x) :=
\Bigl( \frac{1}{\lambda+ 2 \mu} -\frac{1}{\mu} \Bigr) 
\, A_3(\alpha;l;x) .
\ee

The functions (\ref{A1})-(\ref{A3}) can be represented
in terms of some sine and cosine Fourier transforms
in the form
\be \label{A1b}
A_1(\alpha;l;x):= \frac{2}{x} \, S_{1,\alpha} (x, l)
\ee
\be \label{A2b}
A_2(\alpha;l;x) := \frac{4}{x^3} \, S_{3,\alpha} (x, l)
- \frac{4}{x^2} \, C_{2,\alpha} (x, l)
\ee
\be \label{A3b}
A_3(\alpha;l^2;x):= 
-\frac{4}{x} \, S_{1,\alpha} (x, l)
+ \frac{12}{x^3} \, S_{3,\alpha} (x, l)
- \frac{12}{x^2} \, C_{3,\alpha} (x, l) ,
\ee
where $S_{n,\alpha} (x, l)$ and $C_{n,\alpha} (x, l)$
are the sine and cosine Fourier transforms
of corresponding functions in the form
\be
S_{n,\alpha} (x, l) := \int^{\infty}_0 \,
\frac{\sin(k \, x)}{k^n \, 
(1+ l^2_{\alpha} \, k^{\alpha})} \, dk .
\ee
\be
C_{n,\alpha} (x, l) := \int^{\infty}_0 \,
\frac{\cos (k \, x)}{k^n \, 
(1+ l^2_{\alpha} \, k^{\alpha})} \, dk .
\ee
Note that the integrals $C_{n,\alpha} (x, l)$
and $S_{n,\alpha} (x, l)$ do not converge for 
a wide range of parameters.
At the same time, the combinations 
in the form (\ref{A1b})-(\ref{A3b}) converge
for some of these cases. {In this connection, we give}
some values of the function $A_2$ for $x=1$:
\be
A_2(2.0;1;1)= 1.48131675 \quad
A_2(2.7;1;1) = 1.430697955 \quad
A_2(1.2;1;1)= 1.550145537
\ee
\be
A_2(2.0;0.1;1) = 3.078796 \quad
A_2(2.7;0.1;1) = 3.049724200 \quad
A_2(1.2;0.1;1) = 3.098659886
\ee
\be
A_2(2.0;10;1) = 0.20179058 \quad
A_2(2.7;10;1) = 0.3025869013 \quad
A_2(1.2;10;1) = 0.09279446113.
\ee


\newpage

\section{Fractional Gradient Elasticity with Nonlinear Effects}

\subsection{Fractional nonlinear constitutive equations}

Nonlinearity in the fractional gradient elasticity models
can be incorporated in analogy to conventional nonlinear 
stress-strain relations used for `non-gradient' material 
behavior. The theories of nonlinear elasticity and plasticity are
classical examples which have been well developed and used in a 
plethora of applications, by also accounting for thermal, diffusion and
electromagnetic effects. Such type of multiphysics models for deforming
materials lead to systems of differential equations (with derivatives
of integer order) which are not only difficult to address analytically or 
numerically, but also not always able to interpret the observed behavior.
The point of view advanced here is that the multiphysics/multivariable 
approach may be supplemented by a `fractional derivatives' approach utilizing 
a smaller number of variables and phenomenological parameters.
Thus, when a micro/nano structural variable evolves in a complex manner
such that non-fractional micro/nano elasticity or plasticity models cannot 
capture the observed behavior, one may explore the possibility of ignoring the 
explicit
appearance of that variable in the constitutive equations and use non-integer
(instead of integer) derivatives to describe deformation with the newly 
introduced fractional parameter modeling of `missing' phenomenology.

We postpone this general discussion for the future and we focus here on 
extending the linear fractional gradient elasticity {onside} various of the 
previous section to include the nonlinear effects. The proposed nonlinear 
fractional gradient constitutive equation used involves scalar measures of the 
stress and strain tensors; i.e. their second invariants, as these quantities 
enter in both theories of nonlinear elasticity and plasticity.
In plasticity theory, in particular, we employ 
the second invariants of the deviatoric stress 
and plastic strain tensors (\cite{Aifantis1,Aifantis2}). 
The effective (equivalent) stress $\sigma$ 
is defined by the equation
\be 
\sigma = \sqrt{(1/2) \sigma^{\prime}_{ij} \sigma^{\prime}_{ij}} ,
\ee
where $\sigma_{ij}$ is the stress tensor, and
\[ \sigma^{\prime}_{ij} = \sigma^{\prime}_{ij} -\frac{1}{3}\sigma^{\prime}_{kk} \delta_{ij} . \]
The effective (equivalent) plastic strain is defined as
\be
\varepsilon = \int dt \sqrt{2 \dot{\varepsilon_{ij}} 
\dot{\varepsilon_{ij}} } \quad ,
\ee
where $\dot{\varepsilon_{i	j}}$ is the {plastic strain rate tensor, which 
is assumed to be traceless in order to satisfy plastic incompressibility}.
 
{Motivated by the above, we propose the following} form
of nonlinear fractional differential equation {for the scalar quantities 
$\sigma$ and $\varepsilon$ which can be used as a basis for a future tensorial 
formulation of nonlinear elasticity and plasticity theories}
\be \label{eq-01}
\sigma({\bf x}) =
E \, \varepsilon ({\bf x}) + c (\alpha)
 \, ((-\Delta)^{\alpha/2} \varepsilon ) ({\bf x}) +
\eta \, K(\varepsilon ({\bf x})), \quad (\alpha >0),
\ee
where $K(\varepsilon ({\bf x}))$ is a nonlinear function,
which describes the usual (homogenous) 
{material's response beyond linear elasticity or linear hardening 
plasticity};
$c (\alpha)$ is an internal parameter,
that measures the nonlocal character of 
{deformation} mechanisms; 
$E$ is {a shear-like elastic} modulus; 
$\eta$ is a small parameter of non-linearity; and
$(-\Delta)^{\alpha/2}$ is the fractional Laplacian
in the Riesz form \cite{KST}.
As a simple example of the nonlinear function, we can consider 
\be \label{eq-01b}
K(\varepsilon ) = \varepsilon^{\beta} ({\bf x}) ,
\quad (\beta >0) .
\ee
Equation (\ref{eq-01}), where $K(\varepsilon)$ is defined 
by (\ref{eq-01b}) with $\beta=3$,
is the fractional Ginzburg-Landau equation 
(see, for example, \cite{GL-0}--\cite{GL-2}).
{For various choices of the parameters ($E,\eta,\beta$) characterizing the
homogeneous material response, different models of elastic and inelastic 
behavior may result. It is noted, in particular, that (\ref{eq-01}) may be 
considered as a fractional generalization of the flow stress expression for
the conventional theory of plasticity with $E$ denoting the linear hardening 
modulus and $(\eta,\beta)$ the Ludwik--Hollomon parameters.}


Let us derive a particular solution of equation (\ref{eq-01}) with $K(\varepsilon )=0$.
To solve the linear fractional differential equation 
\be \label{FPDE-1b}
\sigma({\bf x}) = E \, \varepsilon ({\bf x}) +
c (\alpha) \, ((-\Delta)^{\alpha/2} \varepsilon ) ({\bf x}) ,
\ee
which represents the constitutive relation
of the fractional gradient elasticity,
we apply the Fourier method and the fractional Green function.


Using Theorem 5.22 of \cite{KST} 
for the case $E \ne 0$ and $\alpha > (n-1)/2$, 
the equation (\ref{eq-03}) is solvable, 
and its particular solution is given by the {expression}
\be \label{Gj-1}
\varepsilon ({\bf x}) = G_{n,\alpha} \, * \, \sigma = 
\int_{\mathbb{R}^n} G_{n,\alpha} ({\bf x} - {\bf x}^{\prime}) \, 
\sigma ({\bf x}^{\prime}) \, d {\bf x}^{\prime} ,
\ee
where the symbol $*$ denotes the convolution operation, and
 $G_{n,\alpha} ({\bf x})$ is defined by (\ref{G-1}),
\be \label{G-1b}
G_{n,\alpha} ({\bf x}) = \frac{|{\bf x}|^{(2-n)/2}}{(2 \pi)^{n/2}} \int^{\infty}_0 
\frac{ \lambda^{n/2} \, J_{(n-2)/2} (\lambda |{\bf x}|) }{c (\alpha) \, \lambda^{\alpha} +E} \, d \lambda,
\ee
where $n=1,2,3$, $\alpha > (n-1)/2$, and 
$J_{(n-2)/2}$ is the Bessel function of the first kind.


Let us consider an unbounded
fractional nonlocal continuum, 
where the stress is applied to an infinitesimally 
small region in {its interior}. 
In this case, we can assume that
the strain $\varepsilon ({\bf x})$ is 
{generated} by a point stress $\sigma({\bf x})$
at the origin of coordinates, i.e. 
\be \label{delta-j}
\sigma({\bf x}) = \sigma_0 \delta({\bf x}) .
\ee
Then, the scalar field $\varepsilon ({\bf x})$ 
{is proportional to the Green's function}, and has the form 
\be \label{phi-Gb}
\varepsilon ({\bf x}) = 
\sigma_0 \, G_{n,\alpha} ({\bf x}) ,
\ee 
i.e. 
\be \label{Pot-2}
\varepsilon ({\bf x}) = 
\frac{1}{2 \pi^2} \frac{\sigma_0}{|{\bf x}|} \, 
\int^{\infty}_0 \frac{ \lambda \, \sin (\lambda |{\bf x}|)}{ 
E + c (\alpha) \, \lambda^{\alpha} } \, d \lambda .
\ee
This is the particular solution of the
fractional-order differential equation (\ref{G-1b}) 
with the point stress for materials distributed
in the three-dimensional space.


\subsection{Perturbation of gradient elasticity by nonlinearity}

Suppose that $\varepsilon ({\bf x})= \varepsilon_0 ({\bf x})$ 
is a solution of equation (\ref{eq-01}) with $\eta=0$, 
i.e. $\varepsilon_0 ({\bf x})$ is a solution of 
the linear {fractional} equation
\be \label{eq-02}
\sigma({\bf x}) =
E \, \varepsilon_0 ({\bf x}) +
c (\alpha) \, ((-\Delta)^{\alpha/2} \varepsilon_0) ({\bf x}) 
.
\ee
The solution of this equation has the form (\ref{Gj-1}). 
We will seek a solution of nonlinear 
equation (\ref{eq-01}) with $\eta \ne 0$
in the form
\be \label{eq-03}
\varepsilon ({\bf x}) = \varepsilon_0 ({\bf x}) + 
\eta \, \varepsilon_1 ({\bf x}) + . . . \ . 
\ee
This means that we consider perturbations to
the strain field $\varepsilon_0 ({\bf x})$ of 
the fractional gradient elasticity,
which are caused by weak {nonlinearity} effects.

In this case, equation (\ref{eq-02}) 
is an approximation of the zero order.
The first order approximation 
with respect to $\eta$ gives the equation
\be \label{eq-04}
E \, \varepsilon_1 ({\bf x}) 
+ c (\alpha) \, ((-\Delta)^{\alpha/2} \varepsilon_1) ({\bf x}) 
+ K (\varepsilon_0({\bf x})) = 0 .
\ee
This equation is equivalent to the linear equation
\be \label{eq-05}
\sigma_{eff}({\bf x}) = E \, \varepsilon_1 ({\bf x}) +
c (\alpha) \, ((-\Delta)^{\alpha/2} \varepsilon_1) ({\bf x}) 
\ee
where the effective stress $\sigma_{eff} ({\bf x})$
is defined by the equation
\be \label{eq-06}
\sigma_{eff} ({\bf x}) = - K (\varepsilon_0({\bf x})) .
\ee
Equation (\ref{eq-04}) can give
a particular solution in the form 
\be \label{Gj-3}
\varepsilon ({\bf x}) = \varepsilon_0 ({\bf x}) + \varepsilon_1 ({\bf x}) =
G_{n,\alpha} \, * \, \sigma + \eta \, G_{n,\alpha} \, * \, \sigma_{eff} ,
\ee
where the symbol $*$ denotes the convolution operation
defined by equation (\ref{Gj-1}). 
Substitution of (\ref{eq-06}) into (\ref{Gj-3}), gives
\be \label{Gj-4}
\varepsilon ({\bf x}) = 
G_{n,\alpha} \, * \, \sigma - \eta \, G_{n,\alpha} \, * \, K( G_{n,\alpha} \, * \, \sigma ) . 
\ee
For point stress (\ref{phi-Gb}), 
equation (\ref{Gj-4}) can written in the form
\be \label{Gj-4b}
\varepsilon ({\bf x}) = 
\sigma_0 \, G_{n,\alpha}({\bf x}) - \eta \, \Bigl( G_{n,\alpha} \, * \, K( \sigma_0 \, G_{n,\alpha} ) \Bigr)({\bf x}) . 
\ee
For the fractional gradient {model} that is described by 
the function $K$ defined by (\ref{eq-01b}), we have
\be \label{Gj-4c}
\varepsilon ({\bf x}) = 
\sigma_0 \, G_{n,\alpha}({\bf x}) - \eta \, \sigma^{\beta}_0 \, 
\Bigl( G_{n,\alpha} \, * \, (G_{n,\alpha})^{\beta} \Bigr) ({\bf x}) , 
\ee
where $\beta>0$.


\subsection{Perturbation by fractional gradient nonlocatity}


Equilibrium value of $\varepsilon_0=const$ 
(where $(-\Delta)^{\alpha/2} \varepsilon_0=0$)
and $\sigma({\bf x})= \sigma=const$ is defined by the condition
\be \label{2eq-01}
E \, \varepsilon_0 + \eta \, K(\varepsilon_0) = \sigma .
\ee
For the case, {where} the function $K$ is defined by Eq. 
(\ref{eq-01b}) with $\beta=3$, we obtain 
the nonlinear algebraic equation
\be \label{2eq-01b}
E \, \varepsilon_0 + \eta \, \varepsilon ^3_0 = \sigma . 
\ee
For $\sigma \ne 0$, there is no solution $\varepsilon_0=0$. 
For $E >0$ and the weak scalar stress field $\sigma\ll \sigma_c$ 
with respect to the critical value 
{$\sigma_c = \sqrt{E / \eta}$}, there exists only one solution
\be \label{2eq-02}
\varepsilon_0 \approx \sigma / E .
\ee

For {negative stiffness materials ($E <0$)} and $\sigma=0$, 
we have three solutions
\be \label{2eq-03}
\varepsilon_0 \approx \pm \sqrt{|E| / \eta } , 
\quad \varepsilon_0 =0 .
\ee
For $\sigma < (2 \sqrt{3}/ 9) \sigma_c$, 
also exist three solutions. 
For $\sigma \gg \sigma_c$, 
we can neglect the first term ($E \approx 0$),
\be \label{2eq-04}
\eta \, \varepsilon ^3_0 \approx \sigma ,
\ee
and {obtain}
\be \label{2eq-05}
\varepsilon_0 \approx (\sigma / \eta)^{1/3} =\sqrt[3]{\sigma/\eta} . 
\ee
In any case, the equilibrium values $\varepsilon_0$ 
are solutions of the algebraic equation (\ref{2eq-01}).

Let us consider a deviation $\varepsilon_1({\bf x})$ 
of the field $\varepsilon ({\bf x})$
from the equilibrium value $\varepsilon_0$.
For this purpose we will seek a solution in the form 
\be \label{2eq-06}
\varepsilon ({\bf x})= \varepsilon_0 + \varepsilon_1 ({\bf x}) .
\ee
In general, the stress field is not constant, i.e. 
$\sigma({\bf x}) \ne \sigma$.
For the first {order} approximation, we have the equation 
\be \label{2eq-07}
\sigma({\bf x}) =
c (\alpha) \, ((-\Delta)^{\alpha/2} \varepsilon_1) ({\bf x}) 
+ \Bigl( E + \eta \, K^{\prime}_{\varepsilon }(\varepsilon_0) \Bigr) \varepsilon_1 ({\bf x}) ,
\ee
where $K^{\prime}_{\varepsilon } = \partial K(\varepsilon ) / \partial \varepsilon $.
Equation (\ref{2eq-07}) is equivalent to the linear fractional differential equation
\be \label{2eq-08}
\sigma({\bf x}) =
E_{eff} \, \varepsilon_1 ({\bf x}) +
c (\alpha) \, ((-\Delta)^{\alpha/2} \varepsilon_1) ({\bf x}) 
\ee
with the effective modulus $E_{eff}$ defined by 
\be
E_{eff} = E + \eta \, K^{\prime}_{\varepsilon }(\varepsilon_0) .
\ee
For the case $K(\varepsilon )=\varepsilon^{\beta}$, 
we have 
\[ E_{eff} = E + \beta \, \eta \, \varepsilon ^{\beta-1}_0 . \]

A particular solution of Eq. (\ref{2eq-08}) 
can be written in the form (\ref{Gj-1}), 
where we use $E_{eff}$ instead of $E$. 
For the point stress (\ref{delta-j}--\ref{Pot-2}), Eq. (\ref{Gj-4}) gives
\be \label{Pot-2b}
\varepsilon_1 ({\bf x}) = 
\frac{1}{2 \pi^2} \frac{\sigma_0}{|{\bf x}|} \, 
\int^{\infty}_0 
\frac{E + E_{eff} + 2 c (\alpha) \, \lambda^{\alpha}}{ 
( c (\alpha) \, \lambda^{\alpha}+ E ) \,
( c (\alpha) \, \lambda^{\alpha}+ E_{eff} ) } 
\, \sin (\lambda |{\bf x}|) \, d \lambda .
\ee
For the case $\alpha=2$, 
the field $\varepsilon_1({\bf x})$ is given by
the equation 
\be
\varepsilon_1 ({\bf x}) = \frac{\sigma_0}{4 \pi c (\alpha) \, |{\bf x}|} \, e^{ - |{\bf x}|/ r_c } ,
\ee
where $r_c$ is defined by 
\be
r^2_c = \frac{c (\alpha)}{ E + \eta \, K^{\prime}_{ \varepsilon }(\varepsilon_0)} .
\ee
It should be noted that on analogous situation exists
in classical theory of electric fields.
In the electrodynamics the field $\varepsilon_1({\bf x})$ describes 
the Coulomb potential with the Debye's screening.
{For the case} $\alpha \ne 2$, 
we have a power-law type of screening 
that is described in \cite{AP2013}.
The electrostatic potential 
for media with power-law spatial dispersion
differs from the Coulomb's potential by the factor
\be 
C_{\alpha,0} (|{\bf x}|) =
\frac{2}{\pi} \, 
\int^{\infty}_0 \frac{ \lambda \, \sin (\lambda |{\bf x}|)}{
E_{eff} + c (\alpha) \, \lambda^{\alpha} } \, d \lambda.
\ee
Note that the Debye's potential differs from the Coulomb's potential by the exponential factor 
$C_D(|{\bf x}|) =\exp (-|{\bf x}|/r_D)$.


\subsection{Fractional Laplacian of Caputo type}

Fractional gradient models can be based on
the Caputo fractional derivatives.
Due to reasons concerning the initial and boundary conditions,
it is more convenient to use the Caputo fractional derivatives.
They allows us to use {simpler} boundary conditions that contain
derivatives of integer order. {In fact}, the initial and boundary
conditions take the same form as for integer-order differential equations.
For fractional derivatives of other type
(for example, the Riemann-Liouville derivatives) 
the boundary conditions are represented by integrals 
and derivatives of non-integer orders \cite{SKM,KST}.

The Caputo fractional derivative is usually denoted
as $ ^CD^{\alpha}_{a+}$, and it is defined by the equation
\begin{equation} \label{Caputo-Diff}
 _a^CD^{\alpha}_x \varepsilon (x) = 
\frac{1}{\Gamma(n-\alpha)} \int^{x}_{a} 
\frac{dx' \, D^n_{x'} \varepsilon (x')}{(x-x')^{1+\alpha-n}}
\end{equation} 
where $n-1<\alpha <n$, and
$\varepsilon (x)$ is a real-valued function defined 
on a closed interval $[a, b]$
such that $\varepsilon (x) \in AC^1[a,b]$ or 
$\varepsilon (x)\in C^1[a,b]$.
We define the left-sided Caputo fractional differential operator
on $[a,b]$ in the form
\be \label{operator-Caputo}
_a^CD^{\alpha}_x[x'] \Bigl( . . . \Bigr) =
\frac{1}{\Gamma(n-\alpha)} \int^x_a \frac{dx'}{(x-x')^{1+\alpha-n}} 
\frac{\partial^n}{\partial {x'}^n} \Bigl( . . . \Bigr)
, \quad (n-1<\alpha<n) .
\ee
The Caputo operator defined by (\ref{operator-Caputo}) 
acts on real-valued 
functions $\varepsilon (x)\in AC^n[a,b]$ as 
$_a^CD^{\alpha}_x[x'] \varepsilon (x')$.
We note that the Caputo operator can be represented as 
\be \label{NotCap} 
_a^CD^{\alpha}_x[x']= _aI^{n-\alpha}_x [x'] D^n[x'] , 
\quad (n-1<\alpha < n) . 
\ee
where we use the left-sided Riemann-Liouville 
fractional integral operator that is defined as
\be \label{operator-RLI}
_aI^{\alpha}_x[x'] \Bigl( . . . \Bigr) 
=\frac{1}{\Gamma(\alpha)} 
\int^x_a \frac{dx'}{(x-x')^{1-\alpha}} \Bigl( . . . \Bigr) , 
\quad (\alpha >0) .
\ee
To designate that the operator given by (\ref{operator-RLI}) 
acts on a real-valued function $\varepsilon (x)\in L_1[a,b]$, 
we employ the notation $_aI^{\alpha}_x[x'] \varepsilon (x')$.

Let us assume that $\varepsilon (x)$ is real-valued functions 
with continuous derivatives up to order $(m-1)$ on 
$W \subset \mathbb{R}^3$, 
such that their $(m-1)$ derivatives are absolutely continuous, 
i.e., $\varepsilon (x) \in AC^m[W]$. 
Then, using the notation introduced in (\ref{NotCap}),
we can define the fractional Laplacian of the Caputo type
\be \label{Delta-C}
^C\Delta^{\alpha}_W = \sum^3_{l=1}( ^CD^{\alpha}_W [x_l])^2.
\ee
In the general case, we have the inequality
$(\, ^C\Delta^{\alpha}_W)^2 \ne \, ^C\Delta^{2\alpha}_W$,
since $(\, ^CD^{\alpha}_x)^2 \ne \, ^CD^{2\alpha}_x$.

In order to solve the corresponding governing fractional equations,
we can use an explicit form of the relationship 
between the square of the Caputo derivative 
$(\, ^CD^{\alpha}_{a+})^2$ and 
the Caputo derivative $\, ^CD^{2\alpha}_{a+}$ of the form
\be \label{Ca2}
(\, ^CD^{\alpha}_{a+})^2 \varepsilon (x) =
\, ^CD^{2\alpha}_{a+} \varepsilon (x) +
\frac{\varepsilon^{\prime}(a)}{\Gamma(1-2\alpha)} 
(x-a)^{1-2\alpha} , \quad (0<\alpha \le 1) ,
\ee
where $\alpha \ne 1/2$.
Using (\ref{Ca2}), we can represent the fractional Laplacian of Caputo 
type as
\be \label{2Lap-Lap2}
\, ^C\Delta^{\alpha}_W \varepsilon ({\bf x})
= \sum^3_{k=1} \, ^CD^{2\alpha}_{x_i} \varepsilon ({\bf x})
+ \sum^3_{k=1} 
\frac{(x_k-a_k)^{1-2\alpha} }{\Gamma(1-2\alpha)} 
\left( \frac{\partial \varepsilon ({\bf x})}{\partial x_k}\right)_{x_k=a_k}.
\ee
Note that the relation given by Eq. (\ref{Ca2}) 
cannot be used for $\alpha>1$.


To describe {nonlinear effects within the aforementioned fractional gradient 
framework}, we consider the constitutive relation in the form
\be \label{3eq-02}
\sigma({\bf x}) = c (\alpha)
 \, ( \, ^C\Delta^{\alpha}_W \varepsilon ) ({\bf x}) +
 K[{\bf x},\varepsilon ({\bf x})] \quad (\alpha >0),
\ee
where we use the fractional Laplacian 
of the Caputo type (\ref{Delta-C})
instead of the Laplacian of the Riesz type.



Let us consider {the situation for} $\sigma({\bf x})=0$ 
for the one-dimensional case 
(${\bf x} \in \mathbb{R}^1$).
In this case, we have
the nonlinear differential equation of order $\alpha >0$
in the form
\be \label{351}
(\, ^CD^{\alpha}_{a+} \varepsilon )(x) = K[x,\varepsilon (x)] \quad 
(\alpha > 0, \quad a \le x \le b), 
\ee
involving the Caputo fractional derivative 
$^CD^{\alpha}_{a+}$ 
on a finite interval $[a,b]$ of the real axis $\mathbb{R}$, 
with the initial conditions
\be \label{352}
\varepsilon^{(k)}(a) = b_k, \quad b_k \in \mathbb{R}, 
\quad (k=0,1,2, . . ., n-1 ) .
\ee
In \cite{KST}, 
the conditions are given for a unique solution $\varepsilon (x)$ 
to this problem in the space $C^{\alpha,r}_{\gamma}[a, b]$ 
defined for $0 <\alpha < n$, $r \in \mathbb{N}$ and 
$0\le \gamma < 1$,  by
$C^{\alpha,r}_{\gamma}[a,b] = \{ \varepsilon (x) \in C^r[a,b]: 
\ ^CD^{\alpha}_{a+} \in C_{\gamma}[a,b]\}$.
One possible way of 
to proceed based on reducing the problem 
considered to the Volterra integral equation \cite{KST}:
\be
\varepsilon (x) =\sum^{n-1}_{j=0} \frac{b_j}{j!} (x-a)^j +
\frac{1}{\Gamma (\alpha)} 
\int^x_a \frac{ K[z,\varepsilon (z)] dz}{(x-z)^{1-\alpha}} , 
\quad (a \le x \le b) .
\ee


In this connection, noted that such equations
with fractional Laplacian of the Caputo type
can be numerically solved.
In \cite{FS2001}, the Cauchy problem of the form 
\be \label{4423}
(\, ^CD^{\alpha}_{0+}\varepsilon ) (x) = K[x, \varepsilon (x)] 
,\quad (x > 0); \quad \varepsilon^{(k)}(0) = b_k ,
\quad (k = 0,1, . . . , n - 1) 
\ee
with the Caputo derivative $\, ^CD^{\alpha}_{0+}$ 
of order $n - 1 < \alpha < n$ ($n \in \mathbb{N}$), 
is numerically solved by using the fixed memory principle 
described in Chapter 8 of \cite{Podlubny}.
In \cite{DFF2004}, an algorithm for the numerical solution
of (\ref{4423}), which is a generalization of 
the classical one-step Adams-Bashforth-Moulton scheme 
for first-order equations, has been suggested.


\newpage

\section{Towards Fractal Gradient Elasticity}



\subsection{Laplacian for non-integer dimensional space}


Let us {begin by giving expressions of fractal--type} differential
operators for functions ${\bf u}=u_r(r)\, {\bf e}_r$
and $\varphi = \varphi(r)$
in the spherical coordinates in $\mathbb{R}^n$ 
for arbitrary $n$ (i.e. $n$ with integer or non-integer values $D$).

As a result, we have equations of  
differential operators in $\mathbb{R}^n$ for continuation
from integer $n$ to arbitrary non-integer $D$ 
in the following forms.

The scalar Laplacian in non-integer dimensional space 
for the scalar field $\varphi=\varphi (r)$ is
\be \label{S-Delta-D}
^S\Delta^{D}_r \varphi= 
\operatorname{Div}^{D}_r \operatorname{Grad}^{D}_{r} \varphi =
\frac{\partial^2 \varphi}{\partial r^2} + \frac{D-1}{r} \, 
\frac{\partial \varphi}{\partial r} .
\ee

The vector Laplacian in non-integer dimensional space 
for the vector field ${\bf u}=u(r) \, {\bf e}_r$ is
\be \label{V-Delta-D}
^V\Delta^{D}_r {\bf u} = 
\operatorname{Grad}^{D}_r \operatorname{Div}^{D}_{r} {\bf u} =
\Bigl(
\frac{\partial^2 u_r}{\partial r^2} + \frac{D-1}{r} \, 
\frac{\partial u_r}{\partial r}  -  \frac{D-1}{r^2} \, u_r 
\Bigr) \, {\bf e}_r.
\ee


Let us consider a case of 
axial symmetry for fractal materials, where
the fields $\varphi(r)$ and ${\bf u}(r)=u_r(r) \, {\bf e}_r$
are also axially symmetric.
Let the $z$--axis be directed along the axis of symmetry. 
Therefore it is convenient to use 
a cylindrical coordinate system.
Equations for differential vector operations
for cylindrical symmetry case have the following forms.

The scalar Laplacian in non-integer dimensional space 
for the scalar field $\varphi=\varphi (r)$ is
\be \label{S-Delta-DC}
^S\Delta^{D}_r \varphi= 
\operatorname{Div}^{D}_r \operatorname{Grad}^{D}_{r} \varphi =
\frac{\partial^2 \varphi}{\partial r^2} + \frac{D-2}{r} \, 
\frac{\partial \varphi}{\partial r} .
\ee

The vector Laplacian in non-integer dimensional space 
for the vector field ${\bf u}=u(r) \, {\bf e}_r$ is
\be \label{V-Delta-DC}
^V\Delta^{D}_r {\bf u} = 
\operatorname{Grad}^{D}_r \operatorname{Div}^{D}_{r} {\bf u} =
\Bigl(
\frac{\partial^2 u_r}{\partial r^2} + \frac{D-2}{r} \, 
\frac{\partial u_r}{\partial r}  -  \frac{D-2}{r^2} \, u_r 
\Bigr) \, {\bf e}_r.
\ee

Using the analytic continuation of Gaussian integrals
the scalar Laplace operator for non-integer dimensional space 
has been already suggested. Specifically
for a function $ \varphi = \varphi (r,\theta)$ 
of radial distance $r$ and related angle $\theta$ 
measured relative to an axis passing through the origin,
the Laplacian in non-integer dimensional space
proposed by Stillinger \cite{Stillinger} is
\be \label{NI-1}
^{St}\Delta^{D} = \frac{1}{r^{D-1}} \frac{\partial}{\partial r} \left( r^{D-1} \,\frac{\partial}{\partial r} \right) +
\frac{1}{r^2 \, \sin^{D-2} \theta} 
\frac{\partial}{\partial \theta} 
\left(  \sin^{D-2} \theta 
\frac{\partial}{\partial \theta} \right) ,
\ee
where $D$ is the dimension of space ($0<D <3$),
and the variables $r \ge 0$, $0\le \theta \le \pi$.
Note that
$(\, ^{St}\Delta^{D} )^2 \ne \, ^{St}\Delta^{2D}$.
If the function depends on the radial coordinate $r$ 
only ($\varphi =\varphi (r)$), then
\be \label{NI-R}
^{St}\Delta^{D} \varphi (r)= \frac{1}{r^{D-1}} \frac{\partial}{\partial r} \left( r^{D-1} \,\frac{\partial \varphi (r)}{\partial r} \right) =
\frac{\partial^2 \varphi (r)}{\partial r^2} +
\frac{D-1}{r} \, \frac{\partial \varphi (r)}{\partial r} .
\ee
It is easy to see that 
the Stillinger's form of Laplacian $\, ^{St}\Delta^{D}$
for radial scalar functions $\varphi ({\bf r})=\varphi (r)$
coincides with the {above} suggested 
scalar Laplacian $\, ^S\Delta^{D}_r$ for this function,
\be \label{NI-R-2}
^{St}\Delta^{D} \varphi (r) = \, ^S\Delta^{D} \varphi (r) .
\ee
The Stillinger's Laplacian can be applied 
only for scalar fields and it cannot be used 
to describe vector fields ${\bf u}=u_r(r) \, {\bf e}_r$ 
because Stillinger's Laplacian for $D=3$ is not equal to 
the usual vector Laplacian for $\mathbb{R}^3$,
\be
^{St}\Delta^{3} {\bf u}(r) \ne \, \Delta {\bf u}(r) =
\Bigl( 
\frac{\partial^2 u_r}{\partial r^2} + \frac{2}{r} \, 
\frac{\partial u_r}{\partial r}  -  \frac{2}{r^2} \, u_r 
\Bigr) \, {\bf e}_r  .
\ee
For the vector fields ${\bf u}=u_r(r) \, {\bf e}_r$, 
we should use the vector Laplace operators 
(\ref{V-Delta-D}) and (\ref{V-Delta-D}).

Note that the gradient, divergence, curl operators and 
vector Laplacian are not considered in Stillinger's \cite{Stillinger}

In \cite{CNSNS2015},
the scalar and vector Laplace operators 
for the case $d \ne D-1$ and
the fields $\varphi=\varphi (r)$ and ${\bf u}=u(r) \, {\bf e}_r$,
are defined by the equations
\be
^S\Delta^{D,d}_r \varphi= 
\operatorname{Div}^{D,d}_r 
\operatorname{Grad}^{D,d}_{r} \varphi , \quad
^V\Delta^{D,d}_r {\bf u} = 
\operatorname{Grad}^{D,d}_r 
\operatorname{Div}^{D,d}_{r} {\bf u} .
\ee
Then, the scalar Laplacian for $d \ne D-1$ 
for the field $\varphi=\varphi (r)$ is
\be \label{S-Delta-Dd}
^S\Delta^{D,d}_r \varphi= 
\frac{\Gamma ((d+\alpha_r)/2) \, \Gamma (\alpha_r/2)}{ 
\pi^{\alpha_r-1/2} \, \Gamma ((d+1)/2)}
\Bigl(  \frac{1}{r^{2 \alpha_r-2}} \, 
\frac{\partial^2 \varphi}{\partial r^2} + 
\frac{d+1-\alpha_r}{r^{2\alpha_r-1}} \, 
\frac{\partial \varphi}{\partial r} \Bigr) ,
\ee
and the vector Laplacian for $d \ne D-1$  
for the field ${\bf u}=u(r) \, {\bf e}_r$ is
\be \label{V-Delta-Dd}
^V\Delta^{D,d}_r {\bf u} = 
\frac{\Gamma ((d+\alpha_r)/2) \, \Gamma (\alpha_r/2)}{ 
\pi^{\alpha_r-1/2} \, \Gamma ((d+1)/2)}
\Bigl( \frac{1}{r^{2 \alpha_r-2}} \, 
\frac{\partial^2  u_r }{\partial r^2} 
+ \frac{d+1-\alpha_r}{r^{2\alpha_r-1}} \, 
\frac{\partial  u_r }{\partial r} 
- \frac{d - \alpha_r}{r^{2\alpha_r}} \,  u_r 
\Bigr) \, {\bf e}_r .
\ee

The vector differential operators and the Laplacian operator
for non-integer dimensional space \cite{CNSNS2015}
allow us to describe complex fractal materials with 
fractal dimensions $D$ for the interior region and 
$d$ for its boundary surface ($d \ne D-1$).


\subsection{Strain and stress in non-integer dimensional space}


Any deformation can be represented as the sum of a pure shear 
and a hydrostatic {component}.
To do this for fractal materials, we can use the identity
\be \label{dii-D}
\varepsilon_{kl} = \Bigl( \varepsilon_{kl} - \frac{1}{D} \delta_{kl} \, \varepsilon_{ii} \Bigr) + \frac{1}{D} \delta_{kl} \, \varepsilon_{ii} .
\ee
The first term on the right is a pure shear,
since the sum of diagonal terms is zero.
Here we use the equation $\delta_{ii}=D$ for non-integer dimensional space (for details see Property 4 
in Section 4.3 of \cite{Collins}).
The second term is the hydrostatic component. 
For $D=3$, equation (\ref{dii-D}) has the well-known form
\be
\varepsilon_{kl} = \Bigl( \varepsilon_{kl} - \frac{1}{3} \delta_{kl} \, \varepsilon_{ii} \Bigr) + \frac{1}{3} \delta_{kl} \, \varepsilon_{ii} ,
\ee
where $\delta_{ii}=3$ is used.

The stress tensor can then be represented (Hooke's law of classical elasticity) 
as
\be \label{eq182}
\sigma_{kl} = K \, \varepsilon_{ii} \, \delta_{kl}
+ 2 \mu \, \Bigl( \varepsilon_{kl} - \frac{1}{D} \delta_{kl} \, \varepsilon_{ii} \Bigr) , 
\ee
where the bulk modulus $K$ 
is related to the Lame coefficients $(\lambda,\mu)$ by
\be
K= \lambda + \frac{2 \, \mu}{D} .
\ee

Under the hydrostatic compression, 
the stress tensor is
\be
\sigma_{kl} = - p \, \delta_{kl} .
\ee
Hence we have 
\be
\sigma_{kk} =- p \, D .
\ee

{which, in view of} (\ref{eq182}), gives
\be
\sigma_{ii} = (\lambda \, D + 2\, \mu) \, \varepsilon_{ii}  .
\ee


The radial component of the strain tensor is 
\be \label{21a}
\varepsilon_{rr} = \frac{\partial u_r}{\partial r} = 
({\bf e}_r,\operatorname{Grad}^{D}_r u_r) . 
\ee
Using $\operatorname{Div}^D_r$ \cite{CNSNS2015}, 
and the trace of the strain tensor
\be \label{21b}
e (r) = Tr[\varepsilon_{kl}] = \varepsilon_{kk} =
\operatorname{Div}^D_r {\bf u} = 
\frac{\partial u_r}{\partial r} + \frac{D-1}{r} \, u_r \, ,  
\ee
we can consider
\be 
e(r) - \varepsilon_{rr}(r)=
\operatorname{Div}^D_r {\bf u} - ({\bf e}_r,\operatorname{Grad}^{D}_r u_r)
= \frac{D-1}{r} \, u_r \, ,
\ee 
as a sum of the angular diagonal components
in spherical coordinates of the strain tensor.
For $D=3$, we have
the well-known {result}
\be \varepsilon_{\theta \theta} + 
\varepsilon_{\varphi \varphi} =
\frac{2}{r} \, u_r  .\ee
When we consider the fractal medium distributed 
in three-dimensional space 
we can define the effective value of the 
diagonal angular components of the strain tensor, as 
\be \label{Eeff}
\varepsilon^{eff}_{\theta \theta} =
\varepsilon^{eff}_{\varphi \varphi} =
\frac{D-1}{2 r} \, u_r .
\ee


Using (\ref{21a}) and (\ref{21b}), 
the components of the stress tensor $\sigma_{kl}=\sigma_{kl}(r)$ 
in spherical coordinates {are given below, starting from the radial 
component} $\sigma_{rr}$ : 
\be \label{sigma-rr}
\sigma_{rr}(r) = 2 \, \mu \, \varepsilon_{rr}(r) + 
\lambda \, e(r)  =
(2\, \mu + \lambda) \, \frac{\partial u_r}{\partial r} + 
\lambda \, \frac{D-1}{r} \, u_r .
\ee

{To deduce appropriate expressions for the fractal counterpart of}
the diagonal angular components, we first note that  
for $D=3$ in spherical coordinates, they are given by the familiar 
relations
\be
\sigma_{\theta \theta} (r) = 
2 \, \mu \, \varepsilon_{\theta \theta}(r) + 
\lambda \, e(r) , \quad
\sigma_{\varphi \varphi} (r) = 
2 \, \mu \, \varepsilon_{\varphi \varphi}(r) + 
\lambda \, e(r) .
\ee

For the fractal medium distributed in three-dimensional space 
we can define the effective value of 
the diagonal angular components of the stress tensor
\be
\sigma^{eff}_{\theta \theta} (r) = 
2 \, \mu \, \varepsilon^{eff}_{\theta \theta}(r) + 
\lambda \, e(r) ,
\ee
\be
\sigma^{eff}_{\varphi \varphi} (r) = 
2 \, \mu \, \varepsilon^{eff}_{\varphi \varphi}(r) + 
\lambda \, e(r) .
\ee
Using the form for the effective components (\ref{Eeff}), we obtain
\be \label{sigma-eff}
\sigma^{eff}_{\theta \theta} (r) = 
\sigma^{eff}_{\varphi \varphi} (r) = 
\lambda \, \frac{\partial u_r}{\partial r} + 
(\lambda +  \mu) \, \frac{D-1}{r} \, u_r .
\ee
This equation define the diagonal angular components 
of the stress tensor in spherical coordinates.


\subsection{Gradient elasticity model for fractal materials}

The standard linear elastic constitutive relation for the isotropic case 
is the well-known Hooke's law -- given by (\ref{eq182}) with $D=3$ -- written 
here 
in 
the alternative form, as 
\be \label{H-0b}
\sigma_{ij} = \lambda \varepsilon_{kk} \delta_{ij} + 2 \mu \varepsilon_{ij} ,
\ee
where $\lambda$ and $\mu$ are the Lame constants. The corresponding governing 
equation for the displacement vector field ${\bf u}={\bf u}({\bf r},t)$, 
reads
\be \label{EL-1}
\lambda \, \operatorname{grad} \operatorname{div} {\bf u}
 + 2 \mu \, \Delta \, {\bf u} + {\bf f} = \rho \, D^2_t {\bf u} ,
\ee
where ${\bf f}={\bf f}({\bf r},t)$ is an external body force field, and 
$\rho$ denotes the density of the material.\\
If the deformation in the material is described by 
${\bf u}({\bf r},t) = u(r,t) \, {\bf e}_r$, then
equation (\ref{EL-1}) has the form
\be \label{EL-2}
(\lambda  + 2 \mu ) \, \Delta \, {\bf u}(r,t) + {\bf f}(r,t) = 
\rho \, D^2_t {\bf u} (r,t) .
\ee


Using the non-integer dimensional vector 
calculus \cite{CNSNS2015},
{we can now suggest a fractal} generalization of equations (\ref{EL-2}) 
for elastic materials in non-integer dimensional space, where
the displacement vector ${\bf u}={\bf u}(r,t)$, 
does not depend on the angular coordinate. It reads
\be \label{VL-1Dd}
(\lambda+2 \mu ) \, \Bigl( 1 \pm l^2_s(D,d) \, ^V\Delta^{D,d}_r \Bigr) 
\, ^V\Delta^{D,d}_r \,{\bf u} +{\bf f} = \rho \, D^2_t {\bf u} .
\ee
where $^V\Delta^{D,d}_r {\bf u}$ 
is the vector Laplacian for $d \ne D-1$  
for the {spherically symmetric} field ${\bf u}=u(r) \, {\bf e}_r$ 
that is defined by the equation
\be \label{V-Delta-Dd2}
^V\Delta^{D,d}_r {\bf u} = 
\frac{\Gamma ((d+\alpha_r)/2) \, \Gamma (\alpha_r/2)}{ 
\pi^{\alpha_r-1/2} \, \Gamma ((d+1)/2)}
\Bigl( \frac{1}{r^{2 \alpha_r-2}} \, 
\frac{\partial^2  u_r }{\partial r^2} 
+ \frac{d+1-\alpha_r}{r^{2\alpha_r-1}} \, 
\frac{\partial  u_r }{\partial r} 
- \frac{d \, \alpha_r}{r^{2\alpha_r}} \,  u_r 
\Bigr) \, {\bf e}_r .
\ee
with $\alpha_r=D-d$.
The vector differential operator (\ref{V-Delta-Dd2})
allow us to describe complex fractal materials with 
boundary dimension $d \ne D-1$.
{Similarly}, we can consider the cylindrical symmetry case
for $d \ne D-1$.

For fractal materials with $d=D-1$ 
equation (\ref{VL-1Dd}) has the form
\be \label{F-1}
(\lambda+2 \mu ) \, ^V\Delta^{D}_r \,{\bf u}(r,t) +{\bf f}(r,t) = 
\rho \, D^2_t {\bf u} (r,t) ,
\ee
where $^V\Delta^{D}_r$ is defined by (\ref{V-Delta-D}).
In the next sections we consider the case $d=D-1$
for {simplicity}.


\subsection{Fractal gradient elasticity for $d=D-1$
and spherical symmetry}

Let us assume that the displacement vector ${\bf u}$ 
is everywhere radial and it is a function of $r=|{\bf r}|$ 
alone ($u_k=u_k(|{\bf r}|,t)$).
Using a continuum model with non-integer dimensional space,
a fractal generalization of gradient elasticity 
equations for this case, where
the displacement vector ${\bf u}={\bf u}(r,t)$, 
does not depend on the {angular coordinate}, has the form
\be \label{VL-1}
(\lambda+2 \mu ) \, \Bigl( 1 \pm l^2_s(D) \, ^V\Delta^{D}_r \Bigr) 
\, ^V\Delta^{D}_r \,{\bf u} +{\bf f} = \rho \, D^2_t {\bf u} .
\ee
This is the gradient elasticity equation 
for homogenous and isotropic fractal materials with 
the spherical symmetry.
Let us consider equation (\ref{VL-1}) 
for static case ($D^2_t {\bf u}=0$) 
with a minus in front of Laplacian, 
i.e. the GRADELA model for fractal materials
\be \label{VL-2}
(\lambda+2 \mu ) \, \Bigl( 1 - l^2_s(D) \, ^V\Delta^{D}_r \Bigr) 
\, ^V\Delta^{D}_r \,{\bf u} + {\bf f} = 0 .
\ee
We can rewrite this equation as
\be \label{VL-3}
(\, ^V\Delta^{D}_r )^2 \,{\bf u} - 
l^{-2}_s(D) \, ^V\Delta^{D}_r \,{\bf u}  
- (\lambda+2 \mu)^{-1} \, l^{-2}_s(D) \, {\bf f} = 0 .
\ee

For spherical symmetry {and}  $d=D-1$,
the vector Laplacian for non-integer dimensional space
has the form \cite{CNSNS2015}
\be \label{VL-4}
^V\Delta^{D}_r {\bf u}(r) =  \Bigl(
\frac{\partial^2 u_r(r)}{\partial r^2} +
\frac{D-1}{r} \, \frac{\partial u_r(r)}{\partial r} 
- \frac{D-1}{r^2} \, u_r(r) \Bigr) \, {\bf e}_r ,
\ee
where we assume that the displacement vector is
${\bf u}({\bf r})=u_r(r) {\bf e}_r$. 
For non-fractal materials {in the three--dimensional space}, we should use 
equation (\ref{VL-4})
with $D=3$.
Equation (\ref{VL-4}) gives  
\[ (\, ^V\Delta^{D}_r)^2 {\bf u}(r)
= \Bigl( \frac{\partial^4 u_r(r)}{\partial r^4} 
+ \frac{2(D-1)}{r} \, \frac{\partial^3 u_r(r)}{\partial r^3} +
\]
\be \label{VL-5}
+ \frac{(D-1)(D-5)}{r^2} \, \frac{\partial^2 u_r(r)}{\partial r^2} 
- \frac{3(D-1)(D-3)}{r^3} \, \frac{\partial u_r(r)}{\partial r}
+ \frac{3(D-1)(D-3)}{r^4} \, u_r(r) \Bigr) \, {\bf e}_r .
\ee
Substitution of expressions (\ref{VL-4})- \ref{VL-5}) and 
${\bf f}(r)=f(r) \, {\bf e}_r$ into equation (\ref{VL-3}) 
gives
\[ \frac{\partial^4 u_r(r)}{\partial r^4} 
+ \frac{2(D-1)}{r} \, \frac{\partial^3 u_r(r)}{\partial r^3} + \left(\frac{(D-1)(D-5)}{r^2} - 
l^{-2}_s(D) \right) \, \frac{\partial^2 u_r(r)}{\partial r^2} -
\]
\[
- \left( \frac{3(D-1)(D-3)}{r^3} +  
l^{-2}_s(D) \frac{D-1}{r}\right) 
\, \frac{\partial u_r(r)}{\partial r} +
\]
\be \label{EM-9}
+ \left( \frac{3(D-1)(D-3)}{r^4} +  
l^{-2}_s(D) \frac{D-1}{r^2}\right) \, u_r(r) 
- (\lambda+2 \mu)^{-1} \, l^{-2}_s(D) \, f(r) = 0.
\ee
The general solution for the case $f(r)=0$ is
\be
u_r(r) = C_1 \, r + C_2 r^{1-D} - C_3 \, I_I(D,r) -  C_4 \, I_K(D,r) ,
\ee
where $I_I(D,r)$ and $I_K(D,r)$ are the integrals of Bessel functions, as 
follows:
\be \label{IID}
I_I(D,r) = D\, r\, \int dr \, r^{-D-1} \int dr \, r^{D/2+1} I_{D/2}(r/l_s(D)) ,
\ee
\be \label{IKD}
I_K(D,r) = D\, r\, \int dr \, r^{-D-1} \int dr \, r^{D/2+1} K_{D/2}(r/l_s(D)) ,
\ee
where $I_{\alpha}(x)$ and $K_{\alpha}(x)$ are Bessel functions of the first and second kinds.


\subsection{Fractal gradient elasticity for $d=D-1$
and cylindrical symmetry}

For cylindrical symmetry {and} $d=D-1$,
the vector Laplacian for non-integer dimensional space has the form
\cite{CNSNS2015}
\be \label{VL-4c}
^V\Delta^{D}_r {\bf u}(r) =  \Bigl(
\frac{\partial^2 u_r(r)}{\partial r^2} +
\frac{D-2}{r} \, \frac{\partial u_r(r)}{\partial r} 
- \frac{D-2}{r^2} \, u_r(r) \Bigr) \, {\bf e}_r .
\ee
Equation (\ref{VL-4c}) gives  
\[ (\, ^V\Delta^{D}_r)^2 {\bf u}(r)
= \Bigl( \frac{\partial^4 u_r(r)}{\partial r^4} 
+ \frac{2(D-2)}{r} \, \frac{\partial^3 u_r(r)}{\partial r^3} +
\]
\be \label{VL-5c}
+ \frac{(D-2)(D-6)}{r^2} \, \frac{\partial^2 u_r(r)}{\partial r^2} 
- \frac{3(D-2)(D-4)}{r^3} \, \frac{\partial u_r(r)}{\partial r}
+ \frac{3(D-2)(D-4)}{r^4} \, u_r(r) \Bigr) \, {\bf e}_r .
\ee
Substitution of expressions (\ref{VL-4c})- \ref{VL-5c}) and 
${\bf f}(r)=f(r) \, {\bf e}_r$ into equation (\ref{VL-3}) 
gives
\[ \frac{\partial^4 u_r(r)}{\partial r^4} 
+ \frac{2(D-2)}{r} \, \frac{\partial^3 u_r(r)}{\partial r^3} + \left(\frac{(D-2)(D-6)}{r^2} - 
l^{-2}_s(D) \right) \, \frac{\partial^2 u_r(r)}{\partial r^2} -
\]
\[
- \left( \frac{3(D-2)(D-4)}{r^3} +  
l^{-2}_s(D) \frac{D-2}{r}\right) 
\, \frac{\partial u_r(r)}{\partial r} +
\]
\be \label{EM-9c}
+ \left( \frac{3(D-2)(D-4)}{r^4} +  
l^{-2}_s(D) \frac{D-2}{r^2}\right) \, u_r(r) 
- (\lambda+2 \mu)^{-1} \, l^{-2}_s(D) \, f(r) = 0.
\ee
The general solution for the case $f(r)=0$ is
\be
u_r(r) = C_1 \, r + C_2 r^{2-D} - C_3 \, I_I(D-1,r) -  C_4 \, I_K(D-1,r) ,
\ee
where $I_I(D-1,r)$ and $I_K(D-1,r)$ 
are defined by equations (\ref{IID}) and (\ref{IKD}).

\newpage

\section{Conclusions}



We proposed (Section 2) a fractional-order generalization 
of three-dimensional continuum gradient elasticity models 
by assuming weak non-locality of 
power-law type {that results to} constitutive relations 
with fractional Laplacian terms.
A three-dimensional constitutive {equation} for 
fractional gradient elasticity is derived 
from integer-order integral relations for nonlocal {elasticity} 
using a fractional Taylor series {expansion} in the wave-vector space.
The suggested fractional constitutive
relations can be connected with {microscopic atomistic--type models}
with long--range interactions, {formulated}
based on lattice with long-range interactions, 
that can be formulated by using
fractional-order differential and integral operators 
on physical lattices, as suggested in 
\cite{JPA2014}--\cite{CMA2017}.


{Fractional generalizations of gradient models with nonlinearity,
employing fractional order Laplace operators of the Riesz and Caputo types,
are also considered (Section 3) for} 
we consider plasticity of non-local continua
with weak nonlocality of power-law type.
Using the perturbation method,
we obtain corrections to {linearized} constitutive relations
{associated with} weak deviations from elasticity, {as well as with}
fractional gradient non-locality.
We assume that such fractional gradient models can be
described by the fractional variational principle
suggested in \cite{TA2015,PJMI2015}.


{Finally, we propose (Section 4)}
models of gradient elasticity for
fractal materials by using vector the calculus
on non-integer dimensional spaces recently
suggested in \cite{CNSNS2015,JMP2014,CRM2015}.
{We consider applications for fractal gradient elasticity theory
for axially symmetric problems in spherical and cylindrical coordinates.
Although the proposed models may not capture all features of underlying 
material fractality, the suggested vector calculus on non-integer dimensional 
spaces may enable us to derive concrete results to be checked with experiments 
commonly employed for characterization of standard (non-fractal) material 
response. This task is facilitated by the fact that the governing equations 
to be solved for fractal gradient elastic materials in `non-integer' space, are
differential equations with derivatives of `integer' order, as the case of 
non-fractal media. Solutions to typical problems for both cases $d=D-1$ and 
$D\ne d-1$ will be given in the future. These will generalize some results 
already obtained for fractal elasticity \cite{CRM2015}. In \cite{RA1993} an 
operator split method (the Ru--Aifantis theorem) has been used to obtain 
solutions to gradient elasticity in terms of solutions of corresponding 
problems in classical elasticity \cite{AA2011,AifantisILG,AMA2008}. A 
generalization of the Ru--Aifantis operator split method can be used to solve  
boundary value problems for fractal gradient elasticity by using the solution 
for fractal non-gradient elasticity given in \cite{CRM2015}. Such an extension 
of the Ru--Aifantis operator split method to fractional/fractal elasticity has 
been illustrated in \cite{TA2015}. Nonlinear elasticity and plasticity effects 
within a fractal formulation based on non-integer space can also be considered 
in a similar way}.  

\section*{Acknowledgement}

Support of the Ministry of Education and Science of Russian Federation under grant no. 14.Z50.31.0039 is acknowledged.


\end{document}